\documentclass[12pt]{article}
\usepackage{graphicx,amsmath,amssymb,color,bm}
\usepackage{ulem,cite}
\usepackage[section] {placeins}

\begin{document}

\centerline {\Large\textbf {Quantization Phenomena of critical Hamiltonians
in 2D systems}}


\centerline{S. C. Chen$^{1}$,  J. Y. Wu$^{2}$, C. Y. Lin$^{1}$,  and M. F.
Lin$^{1}$ }
\centerline{$^{1}$Department of Physics, National Cheng Kung University,
Tainan, Taiwan 701}
\centerline{$^{2}$Center of General Studies, National
Kaohsiung Marine University, Kaohsiung 811, Taiwan} \vskip0.6 truecm

\noindent This review work addresses the recent advances in solving more
comprehensive Hamiltonians. The generalized tight-binding model is developed
to investigate the feature-rich quantization phenomena in emergent 2D
materials. The mutli-orbital bondings, the spin-orbital interactions, the
various geometric structures, and the external fields are taken into
consideration simultaneously. Specifically, the IV-group layered systems,
black phosphorus and MoS$_{2}$ exhibit the unique magnetic quantization.
This is clearly indicated in three kinds of Landau levels (LLs), the
orbital-, spin- and valley-dependent LL groups, the abnormal LL energy
spectra, and the splitting, crossing and anticrossing behaviors. A detailed
comparison with the effective-mass model is made. Some theoretical
predictions have been confirmed by the experimental measurements.

\vskip0.6 truecm

\newpage

\bigskip

\centerline {\textbf {I. INTRODUCTION}} \bigskip

\bigskip How to solve the Hamiltonian is one of the basic topics in physics
science. It is very interesting to comprehend the diverse quantization
phenomena due to the various Hamiltonians in condensed-matter systems,
especially for the feature-rich magnetic quantization. Such Hamiltonians
possess the complex effects coming from the multi-orbital bondings, the
spin-orbital coupling (SOC), the magnetic field ($\mathbf{B}$=B$_{z}\widehat{%
z}$), the electric field ($\mathbf{E}$=E$_{z}\widehat{z}$), the interlayer
hopping integrals, the number of layers, the stacking configurations, the
curved surfaces, the hybridized structures, and the distinct
dimensionalities. The generalized tight-binding model is developed to
include the critical interactions simultaneously. The quantized energy
spectra and wave functions can be computed very efficiently by the exact
diagonalization method even for a rather large Hamiltonian with complex
matrix elements. This model has been used to make systematic studies on the
three-dimensional (3D) graphites \cite{chang2005,Ho2011,Wang2011,Ho2012,Chen2015,Ho2014}, 2D graphenes \cite{Y.K.Huang2014,J.H.Ho2008,C.Y.Lin2014,Y.H.Lai2008,C.Y.Lin2015,T.N.Do2015}, and 1D graphene nanoribbons \cite{Y.C.Huang2007,Y.C.Huang2009,H.C.Chung2016}. It is further extended to the mainstream layered
materials, e.g., other group-IV systems \cite{S.C.Chen2015,J.Y.Wu2015,J.Y.Wu2014,J.Y.Wu20152}, and MoS$_{2}$ \cite{Y.H.Ho2015,Y.H.Ho2014,HoY.H.mos2}. Moreover, the generalized tight-binding model can directly
combine with the single- and many-particle theories to study the other
essential physical properties, such as, magneto-optical properties \cite{Y.C.Huang2008,H.C.Chung2016,Y.P.Lin2015,R.B.Chen2014,R.B.Chen2012,Y.H.Ho2010,Y.H.Ho20101} and Coulomb excitations \cite{J.Y.Wu2015,J.Y.Wu2014,J.Y.Wu20141,J.Y.Wu2011,Y.P.Lin20151}

On the other hand, the perturbation method is frequently used to investigate
the low-energy electronic states and magnetic quantization. It is very
suitable for the condensed-matter systems with monotonous band structures.
For example, the effective-mass model can deal with the magnetic
quantization in monolayer graphene \cite{YisongZheng2002,S.G.Sharapov2004,V.P.Gusynin2005,M.O.Goerbig2011}, AA- and AB-stacked
few-layer graphenes \cite{C.P.Chang2011,E.McCann2006,MikitoKoshino2011,H.Min2008,S.H.R.Sena2011,E.McCann2013}, monolayer silicene
and germanene \cite{Ezawa2012}, MoS$_{2}$ \cite{M.Tahir2016}, and black
phosphorus \cite{PRodin2014}. This model will become too complex or
cumbersome to magnetically quantize the multi-valley and/or multi-orbital
electronic states, such as, the magnetic quantization for the oscillatory
energy bands in ABC- and AAB-stacked graphenes \cite{C.Y.Lin2014,T.N.Do2015}, the seriously distorted Dirac-cone structure in sliding bilayer
graphenes \cite{Y.K.Huang2014}, the three constant-energy loops due to the
significant sp$^{3}$ bondings in monolayer tinene \cite{S.C.Chen2015}, and
the mixed energy bands in hybridized carbon systems \cite{T.S.Li2008,T.S.Li2010,C.H.Lee2011,C.H.Lee20111,M.H.Lee2015}. Furthermore, it is very
difficult to resolve the complex quantization phenomena in the presence of
the non-uniform or the composite external fields \cite{Y.C.Ou2014,Y.C.Ou2013,Y.C.Ou2011,Y.H.Chiu2010}.

The layered condensed-matter systems have stirred a lot of experimental \cite%
{RadisavljevicB2011,K.Hao2016,C.Zhang2016,HNLi2015,W-THsu2015,J.Qi2015}
and theoretical studies \cite{C.P.Chang2011,C.Y.Lin2015,Y.H.Lai2008,C.Y.Lin2014,T.N.Do2015,Y.K.Huang2014,YisongZheng2002,V.P.Gusynin2005,T.Morimoto2013}, mainly owing to the nano-scaled thickness and
the specific symmetries. They are ideal 2D materials for studying the novel
physical, chemical and material phenomena. Furthermore, such systems have
shown high potentials for future technological applications, e.g.,
nano-electronics \cite{EngelM.2012,KhanF.2016,KumarA.2016,LiMY2015,Yu-TingWang2015,A.Kasry2010,Q.Xiang2012,MCRechtsman2013},
optoelectronics \cite{Koppens,Bonaccorso,Tseng,LiuJ.B.,Deng,Yan,Kocaman,Tassin,Vakil,Deng1,Deng2,Abajo} and energy storage
\cite{Shown,Hsu,Ibram,Baughman,Simon,Stoller,Chan,Wang,Bulusheva,Bissett,LiJ.Y.2016,Sun,Gwon,Gwon1}. Few-layer graphenes have been successfully
synthesized by the distinct experimental methods, such as, mechanical
exfoliation \cite%
{Hattendorf,Novoselov,Jayasena,Cooper,Noroozi,Dobbelin,Majee,Arao,Bracamonte,Song,Dou}%
, electrostatic manipulation of scanning tunneling microscopy (STM) \cite%
{YinL.J.,XuP.2013jjap,XuP.2012,Kurys,XuP.2013sur}, and chemical vapor
deposition \cite{YeS.2016,NorimatsuW.,WarnerJ.H.,BiedermannL.B.,BorysiukJ.,ReinaA.,LiuL.,GomezT.,ZhangX.2014,ZhouZ.,KimK.S.,LiX.,ZhouH.,SuC.Y.,BaeS.,VijayaraghavanR.K.,WuX.Y.,BoscaA.,LeeJ.K.,QueY.}. Four
kinds of typical stacking configurations, AAA \cite{BorysiukJ.,LeeJ.K.},
ABA \cite{BiedermannL.B.,LiuL.,ZhouZ.}, ABC \cite{NorimatsuW.,WarnerJ.H.} and AAB \cite{BiedermannL.B.,QueY.}, are clearly identified in the
experimental measurements. It should be noticed that the STM tip can
generate the continuous changes in the stacking configuration, e.g., the
configuration transformation among the ABA, ABC and AAB stackings \cite{XuP.2012}.

The essential electronic properties of planar graphenes are dominated by the
$2p_{z}$-orbital hybridization, the hexagonal honeycomb symmetry, the
stacking configuration and the number of layers. The main features of
low-lying energy bands are further reflected in the rich magnetic
quantization. The tri-layer AAA, ABA, ABC and AAB stackings, respectively,
have the unusual energy dispersions: (1) the linearly intersecting bands
(the almost isotropic Dirac-cone structures) \cite{C.P.Chang2011,C.Y.Lin2015}, (2) the parabolic bands and the linear bands \cite{Y.H.Lai2008},
(3) the weakly dispersive bands, the sombrero-shaped bands, and the linear
bands \cite{C.Y.Lin2014}; (4) the oscillatory bands, the sombrero-shaped
bands, and the parabolic bands \cite{T.N.Do2015}. Such stacking systems
exhibit the novel Landau levels (LLs), in which the rich magneto-electronic
properties include the diverse B$_{z}$-dependent energy spectra, the
asymmetric energy spectra about the Fermi level (E$_{F}$), the crossing
and/or anti-crossing behaviors, the main and side modes, and the
configuration- and E$_{z}$-created splitting states. Specifically, the
configuration transformation between AA and AB stackings will induce the
thorough destruction of the Dirac-cone structures. The three kinds of LLs,
the well-behaved, perturbed and undefined LLs, are predicted to reveal in
the changes from the linear to the parabolic bands

Few-layer germanene and silicene can be synthesized on distinct substrate
surfaces, e.g., Si on Ag(111), Ir(111) \& ZrB2 surfaces \cite{LiT.,VogtP.,B.Aufray,A.Fleurence,L.Meng}; Ge on Pt(111), Au(111) \& Al(111)
surfaces \cite{L.F.Li,M.Derivaz,Davila}. Germanene and silicene possess
the buckled structures with a mixed $sp^{2}$-$sp^{3}$ bonding rather than a
sp$^{2}$ bonding, since the relatively weak chemical bonding between the
larger atoms cannot maintain a planar structure (Fig. 1(b)). These two
systems have the significant SOC's much stronger than that in graphene. The
SOC's can separate the Dirac-cone structures built from the dominating $%
3p_{z}$ or $4p_{z}$ orbitals; that is, the intrinsic systems are narrow-gap
semiconductors (E$_{g}\sim $45 meV for Ge \& E$_{g}\sim $5 meV for Si) \cite%
{C.C.LiuPRB,C.C.LiuPRL}. Furthermore, the application of a uniform
perpendicular electric field leads to the modulation of energy gap and the
splitting of spin-related configurations \cite{Ezawa2012,M.Ezawa2012,N.D.Drummond2012}. The magneto-electronic properties are greatly enriched by SOC
and E$_{z}$, including the modified B$_{z}$-dependent energy spectra, the
spin-up- and spin-down-dominated states, and the E$_{z}$-generated crossing
and anti-crossing behaviors.

Monolayer tinene is successfully fabricated on a substrate of bismuth
telluride \cite{F.Zhu}, while monolayer Pb system is absent in the
experimental measurements up to now. The theoretical studies show that the
single-layer Sn an Pb systems have rather strong sp$^{3}$ bondings and SOC's
\cite{C.C.LiuPRB,Y.Xu2013}. Apparently, the complex chemical bondings
from ($s,p_{x},p_{y},p_{z}$) orbitals need to be included in the low-energy
model calculations. However, the low-lying electronic structures of
graphene, silicene and germanene are mainly determined by the p$_{z}$
orbitals. The very pronounced mixing effects of multi-orbital bondings and
SOC's can create the $p_{z}$- and ($p_{x},p_{y}$)-dominated energy bands
near E$_{F}$, indicating the existence of the multi-constant-energy loops.
There exist two groups of low-lying LLs, with the different orbital
components, spin configurations, localization centers, state degeneracy, and
B$_{z}$- and E$_{z}$-dependencies. Specially, the LL splitting and
anti-crossing behaviors strongly depend on the type of orbitals and the
external fields. The competitive or cooperative relations among the orbital
hybridizations, SOC, \textbf{B} and \textbf{E} are worthy of detailed
investigations.

The group-V phosphorus possesses several allotropes in which black
phosphorus (BP) is the most stable phase under normal experimental
conditions \cite{PLiu}. Few-layer phosphorene is successfully obtained using
the mechanical cleavage approach \cite{PLi2014,PLiu2014}, liquid exfoliation
\cite{PBrent2014,PYasaei2015,Pkang2015}, and mineralizer-assisted short-way
transport reaction \cite{PLange2007,PNilges2008,PKopf2014}. Specially, the
experimental measurements show that the BP-based field effect transistor has
an on/off ratio of 105 and a carrier mobility at room temperature as high as
103 cm$^{2}/$Vs. BP is expected to play an important role in the
next-generation electronic devices \cite{PLi,PLiu2014}. Phosphorene exhibits
a puckered structure related to the $sp^{3}$ hybridization of ($%
3s,3p_{x},3p_{y},3p_{z}$) orbitals. The deformed hexagonal lattice of
monolayer BP has four atoms \cite{PRudenko}, while the group-IV honeycomb
lattice includes two atoms. The low-lying energy bands are highly
anisotropic, e.g., the linear and parabolic dispersions near E$_{F}$,
respectively, along the $\widehat{k_{x}}$ and $\widehat{k_{y}}$ directions.
The anisotropic behaviors are further reflected in other physical
properties, as verified by recent measurements on optical spectra and
transport properties \cite{PLi,PLow}. BP has a middle energy gap of $~\sim
1.5-2$ eV at the $\Gamma $ point, being quite different from the narrow or
zero gaps of group-IV systems. The low-lying energy dispersions, which are
dominated by $3p_{z}$ orbitals, can be described by a four-band model with
the complicated multi-hopping integrals \cite{PRudenko}. The low-energy
electronic structure is easily tuned by a perpendicular electric field,
e.g., the monotonic increase of E$_{g}$ with E$_{z}$ in monolayer BP, and
the transition from a semiconducting to a gapless system in bilayer BP. \cite%
{PDolui,PLiu2015}. In sharp contrast with the group-IV monolayer systems,
monolayer phosphorene presents the unique LLs, with the asymmetric energy
spectrum about $E_{F}$, the reduced state degeneracy, and the
spin-independent configuration. The important differences mainly come from
the geometric structure, the orbital hybridization, and the SOC. The
magnetic quantization is greatly diversified by the number of layers.

The transition metal dichalcogenide monolayers can be produced by the
micromechanical cleavage \cite{K.S.Novoselov2005,Z.Y.Yin2012,H.Li2014,MakK.F.2010}, liquid-phase exfoliation \cite{J.N.Coleman2011,K.G.Zhou2011} and chemical vapor deposition \cite{Y.H.Lee2012,S.Najmaei2013,B.Liu2015,J.C.Shaw2014}. Due to the unusual electronic and
optical properties, various technological applications have been proposed
for these materials, such as, electronic \cite{RadisavljevicB2011,Y.Yoon2011,H.Wang2012,Y.J.Zhang2012,Q.H.Wang2012} and optoelectric \cite%
{Q.H.Wang2012,D.Xiao2012,Y.J.Zhang2012PRB} devices. The high potentials
in the field-effect transistors are supported by the room-temperature
carrier mobility over 200 cm$^{2}/$Vs and the high on/off ratio of $\sim
10^{8}$ \cite{RadisavljevicB2011}. Furthermore, the experimental
measurements show that they have a direct band gap in the visible frequency
range \cite{MakK.F.2010,A.Splendiani2010,J.S.Ross2013} and the
valley-dependent optical selection rules \cite{D.Xiao2012,T.Cao2012}. The
stronger SOC and the inversion symmetry breaking lead to the spin- and
valley-dependent electronic states \cite{D.Xiao2012}. The MoS$_{2}$-related
systems are very suitable for investigating the spintronics and
valleytronics. Specifically, the single-layer MoS$_{2}$ is composed of
staggered honeycomb-like lattice structures in which a single layer of Mo
atoms is sandwiched by two sulfur layers. This semiconducting system has a
direct energy gap of $\sim $1.59 eV \cite{HoY.H.mos2}. The low-lying
electronic states near three valleys centered at the ($K,K^{^{\prime }}$)
and $\Gamma $ points are dominated by the ($%
4d_{z^{2}},4d_{xy},4d_{x^{2}-y^{2}}$) orbitals of Mo atoms. The SOC can
effectively destroy the spin degeneracy of energy bands, especially for the
valence one contributed by the $4d_{xy}$ and $4d_{x^{2}-y^{2}}$ orbitals.
The quantized LLs are characterized by the dominating orbitals and spin
configurations, being enriched by the constant-energy loops in three
valleys. The degeneracy of the $K$ and $K^{^{\prime }}$ valleys is further
lifted\ by \textbf{B,} owing to the cooperation of the site-energy
difference and the magnetic quantization \cite{HoY.H.mos2}.

\bigskip \bigskip \centerline {\textbf {II. GRAPHENE}}

The Hamiltonian of the layered graphene, which is built from the $2p_{z}$%
-orbital tight-binding functions in a unit cell, is expressed as

$H=\underset{\left\langle ij\right\rangle \left\langle ll^{\prime
}\right\rangle }{\sum }-\gamma _{ij}^{ll^{\prime }}C_{il}^{+}C_{jl^{\prime
}}^{{}},\qquad (1)$ \newline
where $\gamma _{ij}^{ll^{\prime }}$ is the intralayer or interlayer hopping
integral, $i$ the lattice site, and $l$ the layer index. $C_{il}^{+}$ ($%
C_{jl^{\prime }}^{{}}$) can create (annihilate) an electron at the $i$-th ($%
j $-th) site of the $l$-th ($l^{\prime }$-th) layer. A hexagonal unit cell
has 2N carbon atoms for a N-layer graphene. Under a uniform perpendicular
magnetic field, there are $4NR_{B}$ carbon atoms in an enlarged rectangular
unit cell (Fig. 1(a)), since the vector potential ($\mathbf{A}=[0,B_{z}x,0]$%
) can induce a periodical Peierls phase. $R_{B}$ is the ratio between flux
quantum ($\phi _{0}$=hc/e) and magnetic flux through a hexagon ($\phi $=3$%
\sqrt{3}$b$^{2}$B$_{z}$/2; b the C-C bond length), e.g., $R_{B}$=2$\times $10%
$^{3}$ at B$_{z}$=40 T. The quantized LLs are highly degenerate in the
reduced first Brillouin zone with an area 4$\pi ^{2}\diagup $3$\sqrt{3}$b$%
^{2}R_{B}$. The (k$_{x}$=$0,$k$_{y}$=$0$) Hamiltonian, with the real matrix
elements, is sufficient in calculating energy spectra and wavefunctions.
Each LL wavefunction is the superposition of the $4NR_{B}$ tight-binding
functions:

$\left\vert \Psi _{\mathbf{k}}\right\rangle =\underset{i,l}{\overset{}{\sum }%
}A_{i}^{l}\left\vert A_{i\mathbf{k}}^{l}\right\rangle +B_{i}^{l}\left\vert
B_{i\mathbf{k}}^{l}\right\rangle .\qquad (2)$ \newline
$A_{i}^{l}$ and $B_{i}^{l}$ are the probability amplitudes of the
subenvelope functions in the two equivalent sublattices.

The hexagonal symmetry in monolayer graphene can create the low-lying
isotropic Dirac-cone structure and thus the well-behaved LLs with the
specific dependence on quantum number ($n^{c,v}$) and field strength. As to
each (k$_{x}$,k$_{y}$) state, all the LLs have eight-fold degeneracy. This
comes from the equivalent $K$ and $K^{^{\prime }}$ valley, the symmetry of $%
\pm \mathbf{B}$ and the spin degree of freedom. At (k$_{x}$=0,k$_{y}$=0),
the state probabilities of the degenerate LLs are localized at the 1/6, 2/6,
4/6 and 5/6 positions of the enlarged unit cell. The (2/6,5/6) and (1/6,4/6)
states, respectively, correspond to the magnetic quantization from the $K$
and $K^{^{\prime }}$ valleys \cite{Y.K.Huang2014}. The 2/6 localized LL
wavefunctions, as shown in Fig. 2, have the normal probability
distributions, being identical to those of a harmonic oscillator. Quantum
number of each LL is characterized by the number of zero points in the
dominating B sublattice. The $n^{c,v}$=0 LLs only come from the B
sublattice. In general, the $n^{c,v}$ LL wavefunctions in the B sublattice
are proportional to the ($n^{c,v}$+1) LL wavefunction in the A sublattice,
directly reflecting the honeycomb symmetry. The same features are revealed
in the 1/6 case under the interchange of two sublattices. Specifically, the
low energy spectrum is characterized by E$^{c,v}$=$\pm $v$_{F}\sqrt{2\hbar
en^{c,v}B_{z}/c}$ (v$_{F}$ the Fermi velocity), consistent with that
obtained from the effective-mass model \cite{YisongZheng2002,V.P.Gusynin2005}. The square-root dependence is suitable at $\left\vert \text{E}%
^{c,v}\right\vert $$<$1 eV, since the linear bands gradually change into the
parabolic bands in the increment of state energy. In addition, the
high-energy LL spectrum can also be obtained by the generalized
tight-binding model \cite{J.H.Ho2008,C.Y.Lin2015}. Specifically, the
dispersionless feature of 2D LLs is dramatically changed by the distinct
dimensions, e.g., the 1D quasi-LLs and the 3D Landau subbands (discussed in
conclusion).

The main features of LLs, energy spectra, spatial distribution modes and
state degeneracy, are dramatically changed by the number of layers, the
stacking configurations, and the perpendicular electric field. The LLs in
the layered graphenes might exhibit the asymmetric energy spectra about the
Fermi level, the non-square-root or non-monotonous dependence on $n^{c,v}$
and B$_{z}$, and the crossing or anti-crossing behaviors, mainly owing to
the critical interlayer hopping integrals (Fig. 3(a)). Such interactions can
induce three kinds of LLs with the distinct distribution modes: (1) the
well-behaved LLs in a single mode, (2) the perturbed LLs with a main mode
and side modes, and (3) the undefined LLs composed of many comparable modes
(Fig. 4). The LL degeneracy will be reduced to half, when the z$\rightarrow $%
-z inversion symmetry is destroyed by the perpendicular electric field (Fig.
12(a)) or the specific stacking configuration. For example, the trilayer
AAB-stacked graphene has an obvious splitting LL spectrum with observable
spacings about 10 meV \cite{T.N.Do2015}.

The ABC-stacked tetralayer graphene and the sliding bilayer graphene are
chosen to see the geometry-enriched magnetic quantization. The former has
four groups of LLs, in which the quantum numbers of the first, second, third
and fourth groups (black, red, blue and green curves) are, respectively,
obtained from the dominant (B$^{1}$, B$^{3}$, B$^{2}$, B$^{4}$) sublattices
at the 2/6 center. Apparently, the valence and the conduction LLs are
asymmetric about E$_{F}$ (Fig. 4). The LL energy spectrum exhibits the
diverse B$_{z}$-dependences, indicating the sensitive changes of energy
bands with wave vectors (Fig. 3(b)). In general, the first group of LLs has
the monotonous dependence, i.e., their energies grow with the increasing B$%
_{z}$. However, the four LLs nearest to E$_{F}$, which mainly arise from the
weakly dispersive energy bands dominated by the surface states (black
curves), have distribution widths smaller than 8 meV even at rather high $%
B_{z}$ (Fig. 4(c)). The LLs, which are localized at two outmost graphene
layers, are absent in the AA-, AB- and AAB-stacked graphenes \cite{C.Y.Lin2014,T.N.Do2015}.

Specifically, the second group of LLs exhibit the abnormal $n_{2}^{c,v}$
sequence and the unusual energy spectrum, as seen in the conduction and
valence states. At rather small B$_{z}$, all the LLs have the reverse
ordering of E$^{c}$($n_{2}^{c}$)$<$E$^{c}$($n_{2}^{c}-1$). They are
initiated at a specific energy corresponding to the cusp $K$ point of the
sombrero-shaped energy band (red curves). This clearly illustrates that LLs
are quantized from the electronic states enclosed by the inner
constant-energy loops. With the increase of B$_{z}$, the higher-$n_{2}^{c}$
LLs come to exist in the normal ordering, since they arise from the outer
constant-energy loops related to parabolic dispersions. The completely
normal ordering of E$^{c}$($n_{2}^{c}$)$<$E$^{c}$($n_{2}^{c}$+1) is revealed
only at B$_{z}>$100 T, directly reflecting the fact that the electronic
states under the cusp-shaped energy dispersions are only quantized into the $%
n_{2}^{c}$=0 LLs. The ordering of LLs is mainly determined by the
competitive relation between the area covered by the cusp-shaped energy
dispersions and the B$_{z}$-enhanced state degeneracy [details in Ref. \cite%
{C.Y.Lin2015}].

The novel intragroup anticrossings appear frequently in the non-monotonous
LL spectrum, as seen in the range of 0.29 eV$<$E$^{c}$($n_{2}^{c}$)$<$0.36
eV. In addition to a main mode, the specific interlayer hopping integrals, ($%
\beta _{3}$,$\beta _{2}$,$\beta _{5}$) (Fig. 3(a)), cause the $n_{2}^{c}$
LLs to possess certain side modes with the zero points of $n_{2}^{c}\pm $3I
(I an integer) \cite{C.Y.Lin2014,C.Y.Lin2015,M.Inoue,T.Morimoto2013}.
The lower-$n_{2}^{c}$ perturbed LLs exhibit the distorted spatial
distributions (Figs. 5(j)-5(k)); that is, they significantly deviate from
the monolayer-like single modes (Fig. 2). For example, with the increase of B%
$_{z}$, the wave functions are drastically changed during the anticrossing
of the $n_{2}^{c}$=0 and 3 LLs, as shown in Fig. 5. When the side mode, with
three zero points in the $n_{2}^{c}$=0 LL (or without zero point in the $%
n_{2}^{c}$=3 LL), becomes comparable with their main mode, the same
oscillation modes in these two LLs can prevent the direct crossing.
Apparently, the intragroup LL anticrossings are derived from the magnetic
quantization of the non-monotonous energy bands, e.g., the existence in the
AAB-stacked graphenes and the absence in the AA- and AB-stacked graphenes
\cite{T.N.Do2015}. It should be noticed that the LL anticrossings are also
presented between any two distinct groups at sufficiently high $B_{z}$ and $%
\left\vert E^{c,v}\right\vert $ \cite{C.Y.Lin2014}, i.e., there exist the
intergroup LL anticrossings. Except for the regimes of these anticrossings,
the third and the fourth groups of energy spectra have the normally
continuous B$_{z}$-dependence (blue and green curves in Fig. 4). The similar
anticrossing behaviors are shown in the AB and AAB stackings, but not in the
AA stacking only with the single-mode LLs.

In addition to the well-behaved and the perturbed LLs, the undefiled LLs can
be created during the transformation of stacking configuration. Specially,
the stacking configuration could be changed by the
electrostatic-manipulation STM \cite{XuP.2012,XuP.2013jjap}. When the
configuration of bilayer graphene is transformed from the AA to AB stacking
by the shift along the armchair direction (Fig. 6(a)), two vertical Dirac
cones gradually change into two pairs of parabolic bands. Each Dirac-cone
structure is seriously distorted and thoroughly separated at the critical
shift of $\sim $6b/8 (Fig. 6(b)). It is impossible to get the low-lying
energy bands from the $K$-point expansion, and so does the LL quantization
using the effective-mass model. The $\delta $=6b/8 stacking exhibits the
eight-fold degenerate LLs, being the same with monolayer graphene. However,
this bilayer system has a lot of undefined LLs, as indicated in the unusual B%
$_{z} $-dependent energy spectrum at $\left\vert \text{E}^{c,v}\right\vert >$
0.3 eV (Fig. 6(c)). Each LL in the second group is composed of various zero
points, and the irregular spatial distribution is very sensitive to the
change in field strength. As a result, it displays the significant
anticrossings with all the LLs in the first group.

\centerline {\textbf {III. SILICENE, GERMANENE \& TINENE}}

For the IV-group inorganic layered systems, the sp$^{3}$ orbital bondings
and the SOC's are included in the critical Hamiltonians. In the bases of $%
\left\{ \left\vert p_{z}^{A}\right\rangle ,\left\vert p_{x}^{A}\right\rangle
,\left\vert p_{y}^{A}\right\rangle ,\left\vert s^{A}\right\rangle
,\left\vert p_{z}^{B}\right\rangle ,\left\vert p_{x}^{B}\right\rangle
,\left\vert p_{y}^{B}\right\rangle ,\left\vert s^{B}\right\rangle \right\}
{}\otimes \left\{ \uparrow ,\downarrow \right\} ,$ the nearest-neighbor
Hamiltonian is expressed as

$H=\underset{\left\langle i\right\rangle ,o,m}{\sum }%
E_{o}C_{iom}^{+}C_{iom}^{{}}+\underset{\left\langle i,j\right\rangle
,o,o^{\prime },m}{\sum }\gamma _{oo^{\prime }}^{\mathbf{R}%
_{ij}}C_{iom}^{+}C_{jo^{\prime }m}^{{}}{}$

$\ \ \ \ \ \ \ \ \ \ \ \ \ \ \ \ \ \ \ \ \ \ \ \ \ \ \ \ \ \ +\underset{%
\left\langle i\right\rangle ,p_{\alpha },p_{\beta }^{{}},m,m^{\prime }}{\sum
}\frac{\lambda _{\text{SOC}}}{2}C_{ip_{\alpha }m}^{+}C_{ip_{\beta }m^{\prime
}}^{{}}(-i\epsilon _{\alpha \beta \gamma }\sigma _{mm^{\prime }}^{\gamma
}),\qquad (3)\newline
$where $i(j)$, $o(o^{\prime })$, and $m(m^{\prime })$ stand for the lattice
site, atomic orbital, and spin, respectively. The first and second terms
are, respectively, the site energy (E$_{o}$) and the nearest-neighbor
hopping integral ($\gamma _{oo^{\prime }}^{\mathbf{R}_{ij}}$). The latter is
determined by the type of atomic orbitals, the translation vector $\mathbf{R}%
_{ij}$, and the angle $\theta $ between $\mathbf{R}_{ij}$ and $\widehat{z}$
(Fig. 1(c)). The details of interaction energies are given in Ref. \cite{C.C.LiuPRB}. The last term represents the SOC on the same atom where $\alpha
,\beta $ and $\gamma $, respectively, denote the $x$, $y$ and $z$
components, and $\sigma $ is the Pauli spin matrix. The SOC strength is,
respectively, predicted to be $\lambda _{\text{SOC}}$=0.034, 0.196; 0.8 eV's
for (Si,Ge,Sn) \cite{C.C.LiuPRB}. The SOC between $\left\vert
p_{x}^{{}}\right\rangle $ and $\left\vert p_{y}^{{}}\right\rangle $ can
create the splitting of states with opposite spin configurations, while that
between $\left\vert p_{z}^{{}}\right\rangle $ and $\left\vert
p_{x}^{{}}\right\rangle $ ($\left\vert p_{y}^{{}}\right\rangle $) leads to
the splitting of states and an interchange of spin configurations.
Specially, the magnetic Hamiltonian of monolayer system is a 32$R_{B}\times $%
32$R_{B}$ Hermitian matrix with complex elements.

The SOC, buckled structure and orbital hybridizations in IV-group can induce
the feature-rich energy bands and diversify the quantized LLs. Germanene and
silicene have the similar band structures, in which the low-lying electronic
states mainly come from the $4p_{z}$ and $3p_{z}$ orbitals, respectively. A
small direct energy gap, which corresponds to the slightly separated Dirac
points, is dependent on the strength of SOC. E$_{g}$ is, respectively, 45
meV and 5 meV for Ge and Si systems (inset in Fig. 7(a)). The first pair of
valence and conduction bands have the doubly degenerate states associated
with the spin-down- and spin-up-dominated equivalent configurations. It is
sufficient to only discuss one of both configurations, as shown in Figs.
7(b) and 7(c) for germanene. Near the $K$ ($K^{\prime }$) point, the valence
states are mainly determined by the $\left\vert 4p_{z}^{B};\downarrow
\right\rangle $ and $\left\vert 4p_{z}^{A};\downarrow \right\rangle $ ($%
\left\vert 4p_{z}^{B};\uparrow \right\rangle $ and $\left\vert
4p_{z}^{A};\uparrow \right\rangle $). Their contributions are very sensitive
to the changes of wave vectors along $K\rightarrow $M ($K^{\prime
}\rightarrow \Gamma $) (solid curves in Fig. 7(b)). The similar behaviors
are revealed in the conduction states under the interchange of the A and B
sublattices (Fig. 7(c)). In addition, the ($4p_{x},4p_{y},4s$) orbitals can
make important contributions to the middle-energy states close to the $%
\Gamma $ point.\bigskip \bigskip

The quantized LLs in monolayer germanene (silicene) are characterized by the
subenvelope functions on the A and B sublattices with $sp^{3}$ orbitals and
two spin configurations. All the low-lying LLs in monolayer germanene
(silicene) belong to the well-behaved modes. They are eight-fold degenerate
for each (k$_{x}$,k$_{y}$) state except the four-fold degenerate LLs of $%
n^{c,v}$=0. As to each localization center, there are two subgroups
characterized by the up- and down-dominated configurations, as indicated in
Fig. 8(a)-8(d) for the 2/6 states. The first and the second subgroups,
respectively, have the $n^{c}$=0 conduction LL and the $n^{v}$=0 valence LL
at E$^{c}$=23 meV and E$^{v}$=$-$20 meV. The former and the latter are
caused by the spin-up and spin-down configurations in the dominating B
sublattice, respectively. The other $n^{c,v}\neq $0 LLs in these two
subgroups are doubly degenerate, and their wave functions are identical
under the interchanges of spins and weights of A and B sublattices\textit{.}
There exist certain important differences between germanene and graphene.
Germanene exhibits the significantly splitting $n^{c,v}$=0 LLs with the
partial contributions from the A sublattice. The weight ratio between the A
and B sublattices are quite different for the valence and conduction LLs. In
addition to the dominating $4p_{z}$ orbitals, the contributions due to the ($%
4p_{x},4p_{y},4s$) orbitals are gradually enhanced as $\left\vert \text{E}%
^{c,v}\right\vert $ grows. However, the opposite is true for graphene (Fig.
2).

A perpendicular electric field applied to buckled systems can split energy
bands and even induce the anti-crossing LL spectra. The destruction of the
z=0 mirror symmetry causes one Dirac cone to become two splitting
structures, when the gate voltage between two sublattices (V$_{z}$) grows
from zero. The lower cone structure approaches to the Fermi level, and
energy gap is vanishing at a critical V$_{z}$ where the linearly gapless
Dirac-cone structure is recovered (the inset of Fig. 9(a)). The dependence
of E$_{g}$ on V$_{z}$ is in the cusp form. Another cone structure is always
away from E$_{F}$. The V$_{z}$-dependent cone structures are quantized into
the unusual LL energy spectra (Fig. 9(a)). The K-valley-dependent (or the $%
K^{^{\prime }}$-valley-dependent) LLs are split according to the magnetic
quantization of the lower and higher Dirac structures. The $n^{c,v}>$0 and $%
n^{c,v}$=0 LLs, respectively, have the four-fold and double degeneracy. The
splitting LL energy spectrum, which corresponds to the lower Dirac cone,
exhibits the non-monotonous V$_{z}$-dependence. As a result, the intra-group
LL anticrossings occur frequently in the plentiful LL energy spectrum. In
addition, the V$_{z}$-induced LL splittings and anticrossings are also
presented in the layered graphenes except for the AA-stacked systems \cite%
{S.J.Tsai2012CPL}.

The cooperation of the electric field and spin-orbital coupling can create
the significant probability transfer between the spin-up and spin-down
configurations and thus the frequent intra-group LL anticrossings. For
example, the $n_{K}^{v}$=2 and $n_{K}^{v}$=3 LLs exhibit the dramatic
changes in the spatial distributions within the critical range of 150 meV$<$V%
$_{z}<$350 meV (green and purple triangles in Fig. 9(b) ). At small V$_{z}$%
's, these two $4p_{z}$-dominated LLs have the same quantum modes on the (A$%
_{\uparrow }$,B$_{\uparrow }$) and (A$_{\downarrow }$,B$_{\downarrow }$)
sublattices (Fig. 9(c) and (d)). However, the weight of distinct spin
configurations is very large and small for the former and the latter,
respectively. With the variation of V$_{z}$, the electric field can induce
the probability transfer between A$_{\uparrow }$ and B$_{\uparrow }$ (A$%
_{\downarrow }$ and B$_{\downarrow }$) sublattices. Furthermore, the
intra-atomic SOC of $4p_{z}$ and ($4p_{x},4p_{y}$) orbitals leads to the
significant distribution change on A$_{\uparrow }$ and A$_{\downarrow }$ (B$%
_{\uparrow }$ and B$_{\downarrow }$) sublattices. This means that the latter
two orbitals play an important role in the anti-crossing behaviors, even if
they have small weights. The comparable probability distributions on the
spin-related sublattices are responsible for the anti-crossings of the $%
n_{K}^{v}$ and $n_{K}^{v}$+1 LLs (the $n_{K}^{c}$ and $n_{K}^{c}$+1 LLs). In
addition, the direct crossing from the $n_{K^{^{\prime }}}^{v}$ and $%
n_{K^{^{\prime }}}^{v}$ +1 LLs (the $n_{K^{^{\prime }}}^{c}$ and $%
n_{K^{^{\prime }}}^{c}$+1 LLs) occurs simultaneously.

Tinene has more low-lying energy bands and diverse LLs, compared with
germanene and silicene. A pair of slightly distorted Dirac cones appears
near the K point, and there are parabolic energy bands initiated at the $%
\Gamma $ point (Fig. 10(a)). The latter, which mainly originate from the ($%
5p_{x},5p_{y}$) orbitals, are attributed to the stronger $sp^{3}$ bonding.
State degeneracy at the $\Gamma $ point is further destroyed by the critical
SOC between $5p_{x}$ and $5p_{y}$ orbitals so that one of the parabolic
bands is very close to the Fermi level. The Dirac-cone structure is
quantized into the first group of LLs (the black curves Fig. 11(a)), in
which the main features are similar to those in germanene (Figs. 8(b) and
8(d)), such as, the p$_{z}$-orbital dominance, localization centers, state
degeneracy, spin configurations, quantum modes on the A and B sublattices,
and B$_{z}$-dependence of LL energy spectrum. However, the first group is in
sharp contrast to the second group. Both $5p_{x}$ and $5p_{y}$ orbitals
dominate the second group of LLs and make the almost same contributions (red
and green curves in Fig. 10 (e)). Such LLs only have two equivalent centers
of 1 and 1/2, and they are doubly degenerate (Fig. 10(d)). For each center,
the splitting LLs are characterized by the up- and down-dominated
configurations ($n_{\Gamma ,\uparrow }^{c,v}$ \& $n_{\Gamma ,\downarrow
}^{c,v}$), being attributed to the significant effect of SOC. The spin-split
LL energy spacing is observable, especially for the larger spacing in the
conduction LLs. This spacing grows with the enhanced weight ration of two
spin configurations on the same sublattice. Specifically, the A and B
sublattices present the same quantum modes, since the nearest-neighbor
hopping integrals near the $\Gamma $ point are roughly proportional to the
square of wave vector. However, the hexagonal symmetry can generate the
linear k-dependence in these atomic interactions near the $K$ point. As to
the first group, this accounts for the mode difference of one between the A
and B sublattices.

Tinene exhibits the rich B$_{z}$-dependent energy spectrum and density of
states (DOS). Two group of LLs have the well-behaved modes, as indicated
from the absence of anti-crossings and the existence of intergroup crossings
(Fig. 11(a)). The LL state energies grow with the increase of field strength
except the almost unchanged $n_{K}^{c,v}$=0 ones. As to the first and the
second groups, the B$_{z}$-dependence is presented in the square-root and
the linear forms, respectively (black and blue curves). This directly
reflects the magnetic quantization from the linear and the parabolic energy
dispersions. The spin-split energy spacings in the second group gradually
become large, since the higher field strength creates more localized LL wave
functions and enhances the spin-up or spin-down dominance. That is to say,
the splitting energies are enlarged by the stronger effects of SOC. The main
differences between two groups of energy spectra are further presented in
DOS (Fig. 11(b)). A lot of strong peaks appear in the delta-function-like
symmetric structure, in which their heights are proportional to state
degeneracy. The single- and double-peak structure originate from the first
and the second groups of LLs; furthermore, the former have the larger peak
spacings. The main features of DOS peaks, structure, height, number and
energy, could be verified from the experimental measurements using scanning
tunneling spectroscopy (STS) \cite{LiG.H.2009,MillerD.L.2009,LuicanA.2011,SongY.J.2010,WangW.X.2015}.

The V$_{z}$-dependent LL energy spectra are quite different among the
IV-group layered systems. The splitting LLs cannot survive only in the
AA-stacked graphenes, since the mirror symmetry is preserved even in the
composite magnetic and electric fields \cite{S.J.Tsai2012CPL}. For
monolayer silicene, the splitting energy spacings are very small, and the LL
anti-crossings and crossings are absent (Fig. 12(a)). The weak SOC and the
large v$_{F}$ (the strong energy dispersion) are responsible for the
monotonous V$_{z}$-dependence. However, monolayer germanene and tinene
frequently exhibit intragroup anti-crossings and crossings (Figs. 9(a) and
12(b)), in which two anti-crossing LLs have quantum number difference of $%
\Delta n$=1. Specifically, the latter have the intergroup crossings between
the $5p_{z}$- and ($5p_{x},5p_{y}$)-dominated LLs. The V$_{z}$-induced
intragroup anti-crossings are also observed in the non-AA-stacked graphenes,
while they arise from two LLs with $\Delta n$=3I \cite{C.Y.Lin2014,C.Y.Lin2015,M.Inoue,T.Morimoto2013}. In addition to V$_{z}$, $\Delta n $=1 and
3I are, respectively, determined by the significant SOC's and certain
interlayer hopping integrals.

\centerline {\textbf {IV. BLACK PHOSPHORUS}} Monolayer phosphorene, with a
puckered honeycomb structure, has a rectangular unit cell. There are four
phosphorus atoms, in which two ones are situated at lower and upper
subplanes (Fig. 13). The low-energy band structure is characterized by the $%
3p_{z}$-orbital hybridizations. The few-layer Hamiltonian is expressed as
\begin{equation*}
H=\sum_{\langle ij\rangle \langle ll^{\prime }\rangle }-t_{ij}^{ll^{\prime
}}C_{il}^{+}C_{jl^{\prime }},
\end{equation*}%
where $t_{ij}^{ll^{\prime }}$ represents the five intralayer (Fig. 13(a))
and four interlayer hopping integrals (Fig. 13(b); details in Ref. \cite%
{PRudenko}). For monolayer system, the magnetic Hamiltonian is a $%
4R_{B}\times 4R_{B}$ Hermitian matrix.

The energy bands are greatly enriched by the complicated multi-hopping
integrals. Monolayer phosphorene has a direct gap of $\sim 1.6$ eV near the $%
\Gamma $ point (Fig. 14(a)), being in sharp contrast with that dominated by
the $K$ point in the group-IV systems. The first pair of energy bands
nearest to $E_{F}$ is, respectively, linear and parabolic along $\Gamma X$
and $\Gamma Y$ directions. The valence (conduction) band is due to the
linearly anti-symmetric (symmetric) superposition of the tight-binding
functions on the upper and lower subplanes. As to bilayer phosphorene, two
pairs of low-lying bands have parabolic dispersions, as shown in Fig. 14(b).
The first and second pairs, respectively, correspond to the in-phase and
out-of-phase combinations of two layers.

The quantized LLs in phosphorene are characterized by the subenvelope
functions on the different subplanes and sublayers. They are localized at
the 1/2 and 2/2 positions of the enlarged unit cell, corresponding to the
magnetic quantization at the $\Gamma$ point. The well-behaved spatial
distributions, as shown in Fig. 15(b), are similar to those of monolayer
graphene. The $3p_{z}$-orbit quantization, with spin degree, is four-fold
degenerate for each $(k_{x},k_{y})$ state. This is in sharp contrast to the
eight-fold degeneracy in the group-IV systems, or the double degeneracy of
the spin- and valley-dependent LLs in MoS$_2$. The LL degeneracy depends on
the number of equivalent valleys and the existence of inversion symmetry ($%
z\rightarrow-z$ and $x\rightarrow-x$). There are two groups of valence and
conduction LLs in bilayer phosphorene (the black and red lines in Fig.
15(a)). Both of them differ from each other in the initial energies and
level spacings. The first and second groups, respectively, correspond to the
in-phase and out-of-phase subenvelope functions on sublayers (Fig. 15(c)).

The highly asymmetric energy dispersion leads to the special dependence of
LL energies on ($n^{c,v}$,$B_{z}$), as clearly indicated in Fig. 16. In
monolayer and bilayer phosphorene, the low-lying LL energies cannot be
described by a simple relation with $n^{c,v}B_{z}$, especially for the
higher energy and field strength. This is different from the square-root
dependence in monolayer group-IV systems (Fig. 11) \cite{J.H.Ho2008}, and
the linear dependence in AB-stacked graphene \cite{C.Y.Lin2014,C.Y.Lin2015} and MoS$_{2}$ (Fig. 19). In general, the LL energies grow with the
increment of $B_z$ monotonously. Only the intergroup LL crossings are
revealed in bilayer system (Fig. 16(b)). However, the intragroup and the
intergroup anticrossings are absent, since all the well-behaved LLs are
quantized from the monotonous band structure in the energy-wave-vector space
(Fig. 14).

\bigskip

\centerline {\textbf {V. MoS$_2$}}

\ A MoS$_{2}$ monolayer consists of three centered honeycomb structures, in
which the middle Mo-atom lattice is sandwiched by two S-atom ones (Fig. 17
). From the previous theoretical studies \cite{MoFP,MoFP1,MoFP2}, the
electronic states close to E$_{F}$ are predominantly contributed \ from the (%
$4d_{z^{2}},4d_{xy},4d_{x^{2}-y^{2}}$) orbitals of Mo atoms. The
three-orbital tight-binding model is sufficient to describe the essential
electronic properties. In the bases of $\left\{ \left\vert
4d_{z^{2}}\right\rangle ,\left\vert 4d_{xy}\right\rangle ,\left\vert
4d_{x^{2}-y^{2}}\right\rangle \right\} \otimes \left\{ \uparrow ,\downarrow
\right\} ,$ the Hamiltonian is given by

$H=\underset{\left\langle i\right\rangle ,o,m}{\sum }%
E_{o}C_{iom}^{+}C_{iom}^{{}}+\underset{\left\langle i,j\right\rangle
,o,o^{\prime },m}{\sum }\gamma _{oo^{\prime }}^{\mathbf{R}%
_{ij}}C_{iom}^{+}C_{jo^{\prime }m}^{{}}+\underset{\left\langle
i\right\rangle ,o,o^{\prime },m}{\sum }\frac{\lambda _{soc}}{2}%
C_{iom}^{+}C_{io^{\prime }m}^{{}}(L_{oo^{\prime }}^{z}\sigma
_{mm}^{z}),\qquad (4)$ \newline
where the first, second and third terms are, respectively, the site energy,
the nearest-neighbor hopping integral and the on-site SOC ($\lambda _{soc}$%
=73 meV). These interaction energies could be found in \cite{MoS2TB}. The
site energies are distinct for the $4d_{z^{2}}$ and ($4d_{xy}$,$%
4d_{x^{2}-y^{2}}$) orbitals, and this difference will result in the
valley-dependent LLs. The SOC is only contributed by the z-component angular
momentum ($L^{z}$) and spin moment ($\sigma _{{}}^{z}$). This interaction
occurs between $\left\vert 4d_{x^{2}-y^{2}}\right\rangle $ and $\left\vert
4d_{xy}\right\rangle $ with the same spin configuration, while it is
independent of $\left\vert 4d_{z^{2}}\right\rangle $. As to the magnetic
Hamiltonian, 2$R_{B}$ Mo atoms in an enlarged unit cell can build a 12$%
R_{B}\times $12$R_{B}$ Hermitian matrix.

The multi-orbital bondings and the SOC cause monolayer MoS$_{2}$ to exhibit
the unusual electronic structure. A direct energy gap of 1.59 eV at the $K$
or $K^{^{\prime }}$ point, as shown in Fig. 18(a), is dominated by the site
energies of distinct orbitals. The significant orbital hybridizations lead
to the strong wave-vector dependence. The electronic states of parabolic
bands near E$_{F}$ are centered at the $K$, $K^{^{\prime }}$ and $\Gamma $
points. Furthermore, the SOC can create the spin-split energy bands, e.g.,
the largest splitting energy is 2$\lambda _{soc}$ at the $K$ and $%
K^{^{\prime }}$ points. Whether there exist the splitting spin-up and
spin-down energy bands is dependent on the components of $4d_{x^{2}-y^{2}}$
and $4d_{xy}$ orbitals. The contributions of these two orbitals, as
indicated in Figs. 18(b) and 18(c), are comparable in the splitting valence
bands near the $K$ and $K^{^{\prime }}$ points. However, when the electronic
states mainly come from one of them, or the $4d_{z^{2}}$ orbital, the spin
splitting is very weak, e.g., the lower-lying conduction bands in Fig.
18(d). It is also noticed that the IV-group systems do not have the
spin-split energy bands as a result of the mirror symmetry in A and B
sublattices (Figs. 7(a) and 10(a)).

MoS$_{2}$ systems exhibit the novel magnetic quantization, since the valley-
and spin-dependent LL subgroups could survive simultaneously. All the LLs
have two degenerate localization centers, the 1/2 and 2/2 localization
centers in an enlarged unit cell, e.g., the 1/2 localized LL wavefunctions
shown in Figs. 19(b) and 20(b). The dominating modes have the well-behaved
spatial probability distributions. Each mode is fully determined by the
spin-up or spin-down configuration, but not a superposition of two opposite
spins as revealed in IV-group systems (Figs. 8(b), 8(d) and 10(e)). Each LL
group of monolayer MoS$_{2}$ only corresponds to the occupied LLs or the
unoccupied LLs, while that in IV-group systems includes the valence and
conduction ones. It is further split into LL subgroups under the destruction
of the spin and/or valley degeneracy.

As to the valence LLs, they are magnetically quantized from the electronic
states centered at the $\Gamma $, $K$ and $K^{^{\prime }}$ points (Fig.
19(a)). The $\Gamma $-dependent LL wavefunctions are independent of spin
configuration (Fig. 19(b)), mainly owing to the $d_{z^{2}}$-orbital
dominance and the almost vanishing SOC. The LL energies linearly grow with B$%
_{z}$ (the blue curves in Fig. 19(a)), directly reflecting the parabolic
dispersion near the $\Gamma $ point. The similar B$_{z}$-dependence is
revealed in the energy spectra of other LL subgroups. However, the spin-up
and spin-down (spin-down \& spin-up) LL subgroups, which come from the $K$ ($%
K^{^{\prime }}$) valley, are initiated at $-$0.792 eV and $-$0.938 eV,
respectively. The spin-split LL subgroups are closely related to the $%
4d_{x^{2}-y^{2}}$- and $4d_{xy}$-dominated SOC's, as indicated in Figs.
18(b) and 18(c). Specifically, the degeneracy of two valleys is clearly
destroyed in the increase of B$_{z}$. That is to say, there also exist the $K
$- and $K^{^{\prime }}$-dependent LL subgroups, as shown in Fig. 19(a). The
energy spacing is observable for a sufficiently high B$_{z}$, e.g., $\sim $%
15 meV between the $n_{K,\uparrow }^{v}$=0 and $n_{K^{^{\prime }},\downarrow
}^{v}$ LLs at B$_{z}$=40 T. By the detailed analysis, the site-energy
differences in the B$_{z}$-enlarged unit cell is responsible for these LL
subgroups \cite{HoY.H.mos2}. The coexistence of the spin- and
valley-dependent LL subgroups is absent in IV-group systems. On the other
hand, the lower-lying conduction LLs exhibit the significantly $K$- and $%
K^{^{\prime }}$-dependent subgroups, and the very weak splittings in the
spin-dependent subgroups (Fig. 20).

\bigskip

Up to now, STS has served as a powerful experimental method for
investigating the magneto-electronic energy of the layered graphenes. The
measured tunneling differential conductance (dI/dV) is approximately
proportional to DOS, and it directly reflects the structure, energy, number
and height of the LL peaks. Part of theoretical predictions on the LL energy
spectra are confirmed by STS measurements, such as, the $\sqrt{\text{B}_{z}}$%
-dependent LL energy in monolayer graphene \cite{LiG.H.2009,MillerD.L.2009,LuicanA.2011,SongY.J.2010,WangW.X.2015}, the linear B$_{z}$%
-dependence in AB-stacked bilayer graphene \cite{RutterG.M.2011,YinL.J.2016}, the coexistent square-root and linear B$_{z}$-dependences in
trilayer ABA stacking \cite{YinL.J.2015}, and the 3D and 2D
characteristics of the Landau subbands in AB-stacked graphite \cite{LiG.H.2007}. The other unusual magneto-electronic properties in the layered
systems could be further verified using STS, including the normal and
abnormal B$_{z}$-dependences in ABC-stacked graphenes, three kinds of LLs in
sliding stacking systems, the SOC-induced spin-dominated LLs in germanene
and silicene, two groups of low-lying LLs in tinene, the spin- and
valley-dependent LLs in MoS$_{2}$, the special $n^{c,v} $- and B$_{z}$%
-dependence of LL energies in few-layer phosphorene, and the B$_{z}$- and V$%
_{z}$-dependent energy spectra with the LL splittings, crossings and
anti-crossings. The STS examinations can provide the critical informations
in the lattice symmetry, stacking configuration, SOC, and single- or
multi-orbital hybridization.

\centerline {\textbf {VI. CONCLUDING REMARKS}}\bigskip \bigskip\

In this review work, the generalized tight-binding model, based on the
subenvelope functions of distinct sublattices, is developed to explore the
feature-rich magneto-electronic properties of layered systems. Such model is
suitable for the various symmetric lattices, the multi-layer structures, the
low-symmetry stacking configurations, the distinct dimensions, the
multi-orbital bondings, the coupling interactions of orbital and spin, the
composite external fields, and the uniform and modulated fields \cite{Y.C.Ou2014,Y.C.Ou2013,Y.C.Ou2011,Y.H.Chiu2010}. It is useful in
understanding the essential physical properties, e.g., the diverse
magneto-optical selection rules \cite{Y.C.Ou2014,Y.C.Huang2007,Y.H.Ho20101,R.B.Chen2012,Y.H.Ho2014} and the LL-induced plasmons \cite{J.Y.Wu20141,Chen2015,J.Y.Wu2011}. Moreover, this method could be further used
to solve the new Hamiltonians of the emerging materials under the external
fields.

The layered systems exhibit the unusual energy bands and the rich LLs in
terms of the spatial distributions, orbital components, spin configurations,
state degeneracy, and external-field dependences. The well-behaved,
perturbed and undefined LLs are revealed in sliding graphenes, especially
for the third ones mainly coming from the dramatic transformation between
two high-symmetry stacking configurations. For the ABC- and AAB-stacked
graphenes, the complicated interlayer hopping integrals result in the
abnormal $n^{c,v}$ ordering and the non-monotonic B$_{z}$-dependence. The
intragroup and intergroup LL anti-crossings occur frequently, clearly
illustrating the strong competition of the constant-energy loops in the
magnetic quantization. The SOC can create the up- and down-dominated LLs in
silicene, germanene, tinene and MoS$_{2}$. Tinene and MoS$_{2}$ have the
low-lying LLs composed of the different orbitals and spin configurations,
mainly owing to the cooperation of the critical multi-orbital bondings and
SOC. Concerning few-layer phosporene, the puckered structure induces the
intralayer and the interlayer multi-hopping integrals and thus the special
dependence of LL energies on ($n_{{}}^{c,v}$,B$_{z}$). The LL state
degeneracy is reduced, when the inversion symmetry of z$\longrightarrow -$z
(x$\longrightarrow -$x) is destroyed, or two equivalent valleys are absent.
For example, there are four-fold degenerate LLs in AAB-stacked graphene and
few-layer phosphorene, and doubly degenerate $n_{\Gamma }^{c,v}$ LLs in
tinene. Specifically, MoS$_{2}$ exhibits the spin- and valley-dependent LL
subgroups, in which they arise from the SOC and the site-energy differences
in the B$_{z}$-enlarged cell, respectively. The LL splittings are easily
observed in the presence of a perpendicular electric field except for the
AA-stacked graphenes and monolayer slicene. Furthermore, they induce the
frequent anti-crossings and crossings in the V$_{z}$-dependent energy
spectra. The two anti-crossing LLs, which are, respectively, associated with
the interlayer hopping integrals and the significant SOC, have the quantum
number differences of $\Delta n$=1 and 3I. The above-mentioned diverse LL
energy spectra are directly reflected in the structure, height, energy and
number of the prominent DOS peaks; furthermore, they could be verified by
the STS measurements.

Also, the different dimensions can diversify the magnetic quantization. The
3D graphites and 1D graphene nanoribbons are very different from the 2D
graphenes in the quantized electronic properties. As to graphites, the
periodical interlayer hopping integrals induce the Landau subbands with
energy dispersions along $\widehat{k_{z}}$. The AA-, AB- and ABC-stacked
graphites, respectively, possess one group, two groups, and one group of
valence and conduction Landau subbands, in which the band widths are about 1
eV, 0.2 eV and 0.01 eV \cite{Ho2011,Ho2014,HoC.H.20114938,HoC.H.2013NJP}%
. However, there are N groups of LLs in N-layer graphene systems. The AA-
and ABC-staked graphites exhibit the monolayer-like wave functions, while
the AB-stacked graphite displays the monolayer- and bilayer-like spatial
distributions. In sharp contrast to the ABC-stacked graphenes, the
anti-crossing phenomena in the B$_{z}$-dependent energy spectrum are absent
in the rhombohedral graphite. On the other hand, the magneto-electronic
properties of graphene nanoribbons are mainly determined by the rather
strong competition between the magnetic quantization and the finite-width
quantum confinement \cite{Y.C.Huang2007}. When ribbon widths are larger
than magnetic lengths, 1D nanoribbons have many composite energy subbands.
Each subband is composed of a dispersion-less quasi-LL (QLL) and parabolic
dispersions along $\widehat{k_{x}}$. Such QLLs belong to the well-behaved
modes localized at the ribbon center. Their magneto-optical selection rule
is similar to that of monolayer graphene. In addition, a carbon nanotube
could be regarded as a rolled-up graphene tubule. Except for a super-high
magnetic field (B$_{z}>$10$^{5}$ T; \cite{AjikiH.1996}), it is impossible
to generate the quantized QLLs in a curved cylinder because of the vanishing
net magnetic flux.

\bigskip

\centerline {\textbf {ACKNOWLEDGMENT}}

\bigskip\ This work was supported by the NSC of Taiwan, under Grant No. NSC
102-2112-M-006-007-MY3.

\noindent ~~~~$^{\star }$e-mail address: mflin@mail.ncku.edu.tw

$^{\star }$e-mail address: yarst5@gmail.com 

\bigskip \vskip0.6 truecm

\noindent

\newpage

\centerline {\textbf {Figure Captions}}

Figure 1: (a) Geometric structure of honeycomb graphene with an enlarged
rectangular unit cell in B$_{z}\widehat{z}$, and (b) the buckled (silicene,
germanene,tinene) with (c) the $sp^{3}$ bondings.

Figure 2: For monolayer graphene at B$_{z}$=40 T, (a) the low-lying valence
and conduction LLs, and the probability distributions of the subenvelope
functions at the (b) A and (B) sublattices. The unit of the x-axis is 2R$%
_{B} $, in which i represents the i-th A or B atom in an enlarged unit cell.

Figure 3: (a) Geometric structure and (b) energy bands of the ABC-stacked
tetralayer graphene. $\gamma_i\,^{\prime }s$ denote the intralayer and
interlayer hopping integrals.

Figure 4: The B$_{z}$-dependent LL energy spectrum of the ABC-stacked
tetralayer graphene for (a) the first group, (b) the other three groups; (c)
the four LLs nearest to E$_{F}$.

Figure 5: (a) The intragroup and intergroup LL anticrossing phenomena in the
ABC-stacked tetralayer graphene. The spatial evolutions of subenvelope
functions are shown in (b)--(i) for the second group of LLs, and (j)--(q)
for the second and the third

groups of LLs.

Figure 6: (a) Geometric structure of sliding bilayer graphene along the
armchair direction; (b) energy bands and (c) B$_{z}$-dependent LL spectrum
at $\delta $=6b/8.

Figure 7: (a) Energy bands of monolayer germanene and silicene, and (b)\&(c)
orbital-decomposed state probabilities along the high-symmetry points. The
inset of (a) is the band structure near the Dirac point.

Figure 8: For germanene at B$_{z}$= 15 T, (a) \& (b) the up-dominated and
(c) \&(d) the down-dominated LL energies and spatial probability
distributions, corresponding to the quantized $K$-valley states.

Figure 9: (a) The V$_{z}$-dependent LL energy spectra of germanene at B$_{z}$%
=15 T, and (b) the LL crossing and anticrossing within a certain range of E$%
^{v}$. (c) and (d) the drastic changes of probability distributions during
the LL anticrossings. The inset of (a) is the band structure at B$_{z}$=0
and a critical V$_{z}$. The LL anticrossings are indicated by the red
circles.

Figure 10: (a)-(c) Similar plot as Fig. 7, but shown for monolayer tinene.
Also plotted in (b) are those from the quantized $\Gamma $-valley states.

Figure 11: (a) The B$_{z}$-dependent energy spectra of the first and second
groups (black and blue curves) in tinene, and (b) density of states at B$%
_{z} $ = 30 T.

Figure 12: The V$_{z}$-dependent LL energy spectra at B$_{z}$=15 T for (a)
tinene and (b) silicene.

Figure 13: (a) The geometric structure and the first Brillouin zone of
monolayer phosphorene, and (b) the AB stacking structure of bilayer
phosphorene. Also, the intralayer and interlayer hopping integrals are
marked in (a) and (b), respectively.

Figure 14: Energy bands of (a) monolayer and (b) bilayer phosphorene.

Figure 15: (a) The LL energies of bilayer phosphorene at B$_{z}$=30 T and
(b) the probability distributions. Also shown in (c) are the amplitudes of $%
n^{c}=1$ of the upper (solid curve) and lower (dashed curve) layers.

Figure 16: For monolayer and bilayer phosphorene, the $n^{c}$- and $B_{z}$%
-dependent LL energies are shown in (a) and (b) respectively. The black and
purple dashed lines in (a) and (b) represent the linear dependence.

Figure 17: (a) Geometric structures for MoS$_{2}$ monolayer with an enlarged
rectangular unit cell in B$_{z}\widehat{z}$ and (b) the structure of
trigonal prismatic coordination.

Figure 18. (a) Energy bands of monolayer MoS$_{2}$, and (b)-(d) the
orbital-decomposed state probabilities along the high-symmetry points.

Figure 19. (a) The B$_{z}$-dependent energy spectra of the valence LLs are,
respectively, related to the quantized states near the ($K$,$K^{^{\prime }}$%
) and $\Gamma $ points (black and blue curves), in which the
valley-dependent (spin-dependent) subgroups are represented by the solid and
dashed curves (the red and yellow colors). The spatial probability
distributions are shown in (b) at B$_{z}$=40 T.

Fig. 20. Same plot as Fig. 19, but shown for the low-lying conduction LLs.

\begin{figure}[p]
    \centering
    \includegraphics[width=0.9\textwidth]{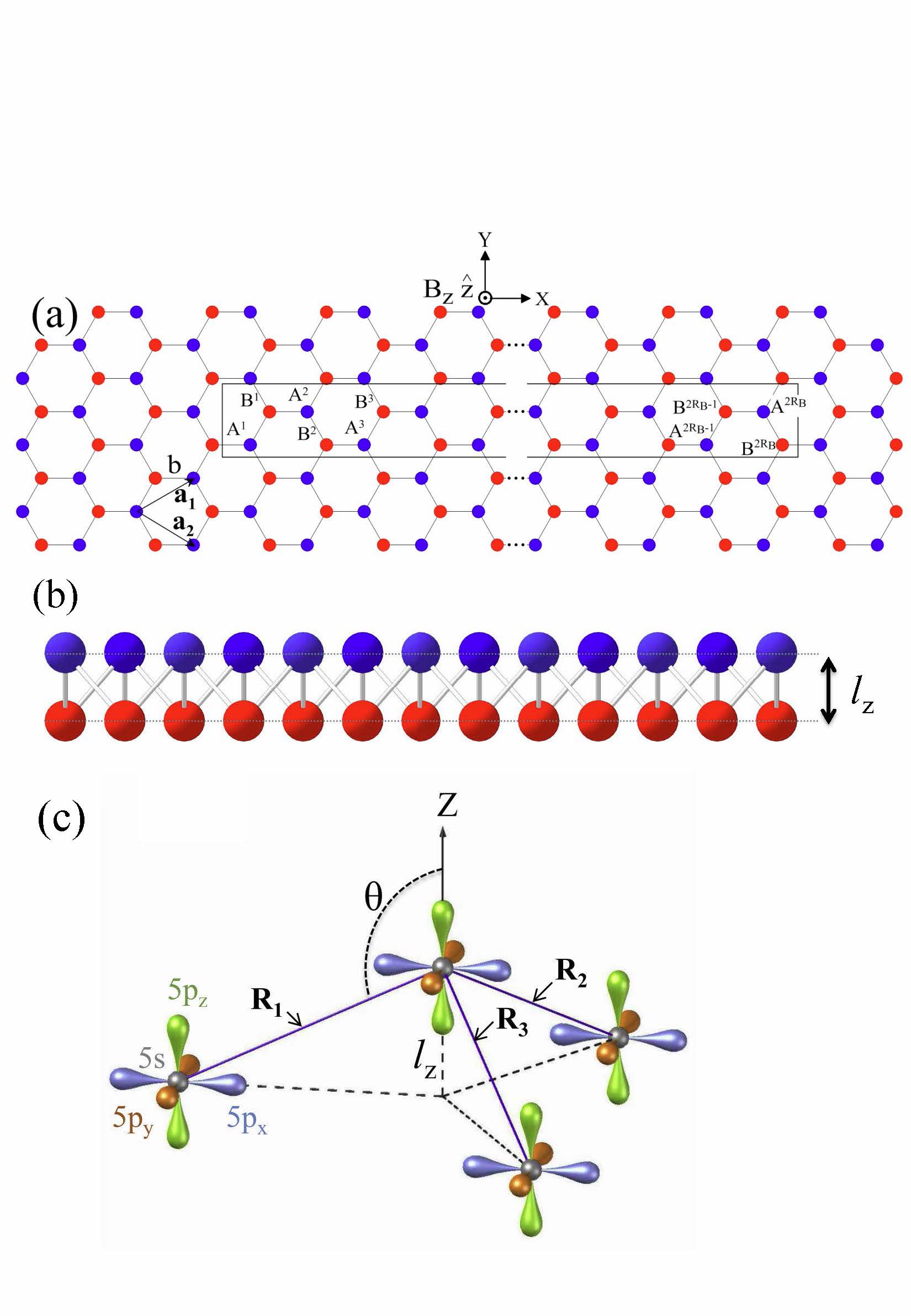}
    \caption{(a) Geometric structure of honeycomb graphene with an enlarged
rectangular unit cell in B$_{z}\widehat{z}$, and (b) the buckled (silicene,
germanene,tinene) with (c) the $sp^{3}$ bondings.}
    \label{figure:1}
\end{figure}

\begin{figure}[p]
    \centering
    \includegraphics[width=0.8\textwidth]{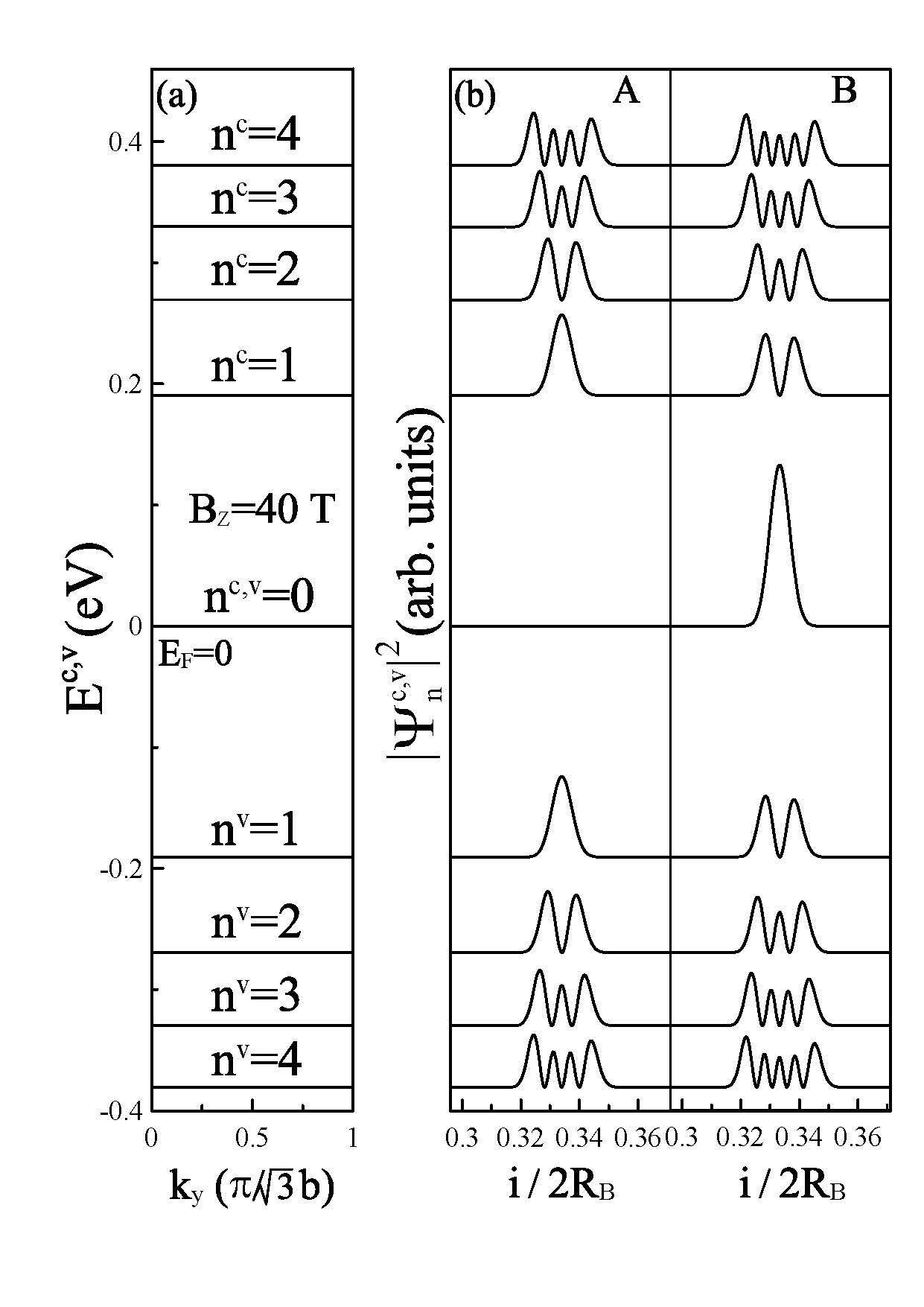}
    \caption{For monolayer graphene at B$_{z}$=40 T, (a) the low-lying valence
and conduction LLs, and the probability distributions of the subenvelope
functions at the (b) A and (B) sublattices. The unit of the x-axis is 2R$%
_{B} $, in which i represents the i-th A or B atom in an enlarged unit cell.}
    \label{figure:2}
\end{figure}

\begin{figure}[p]
    \centering
    \includegraphics[width=0.8\textwidth]{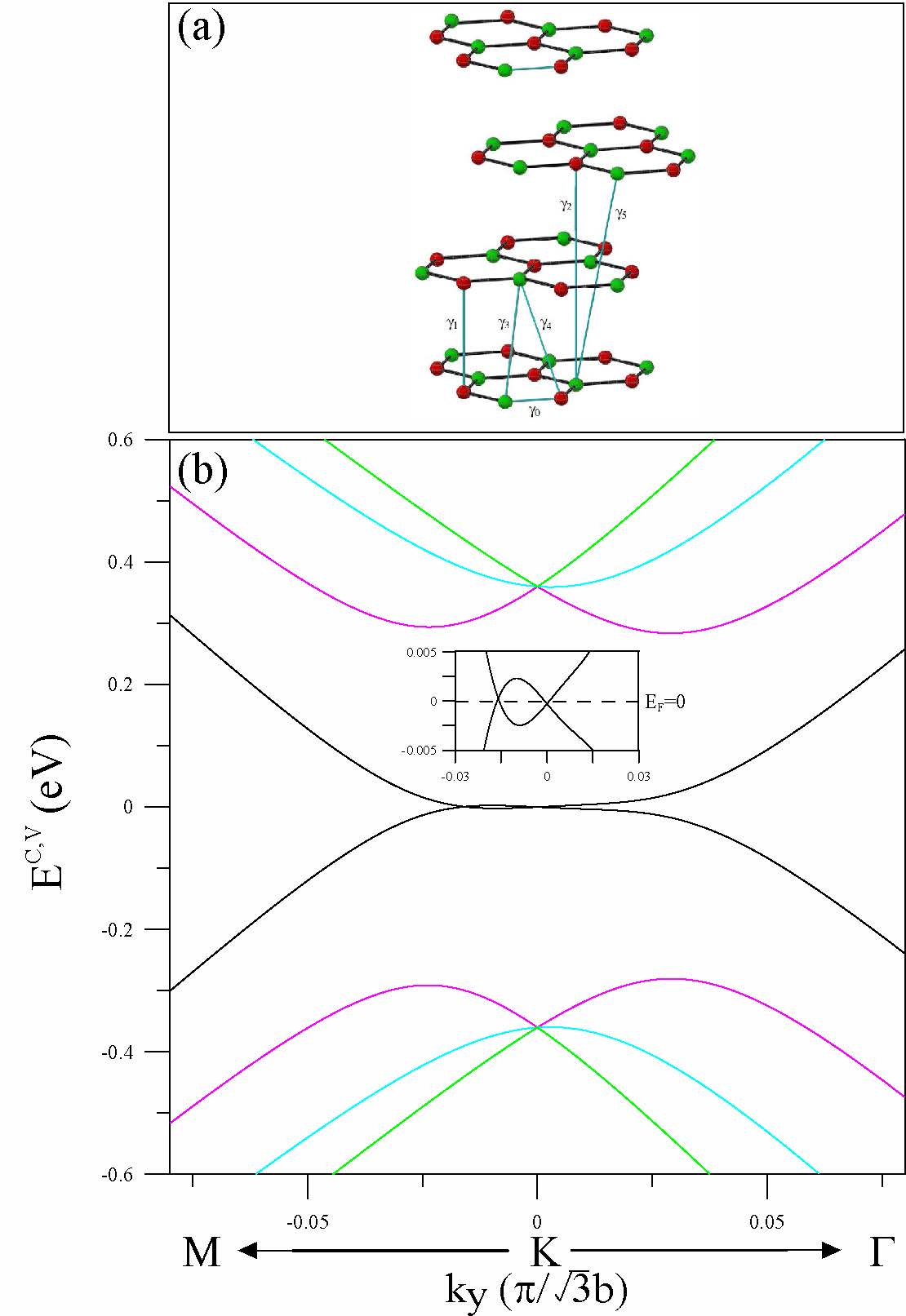}
    \caption{(a) Geometric structure and (b) energy bands of the ABC-stacked
tetralayer graphene. $\gamma_i\,^{\prime }s$ denote the intralayer and
interlayer hopping integrals.}
    \label{figure:3}
\end{figure}

\begin{figure}[p]
    \centering
    \includegraphics[width=0.9\textwidth]{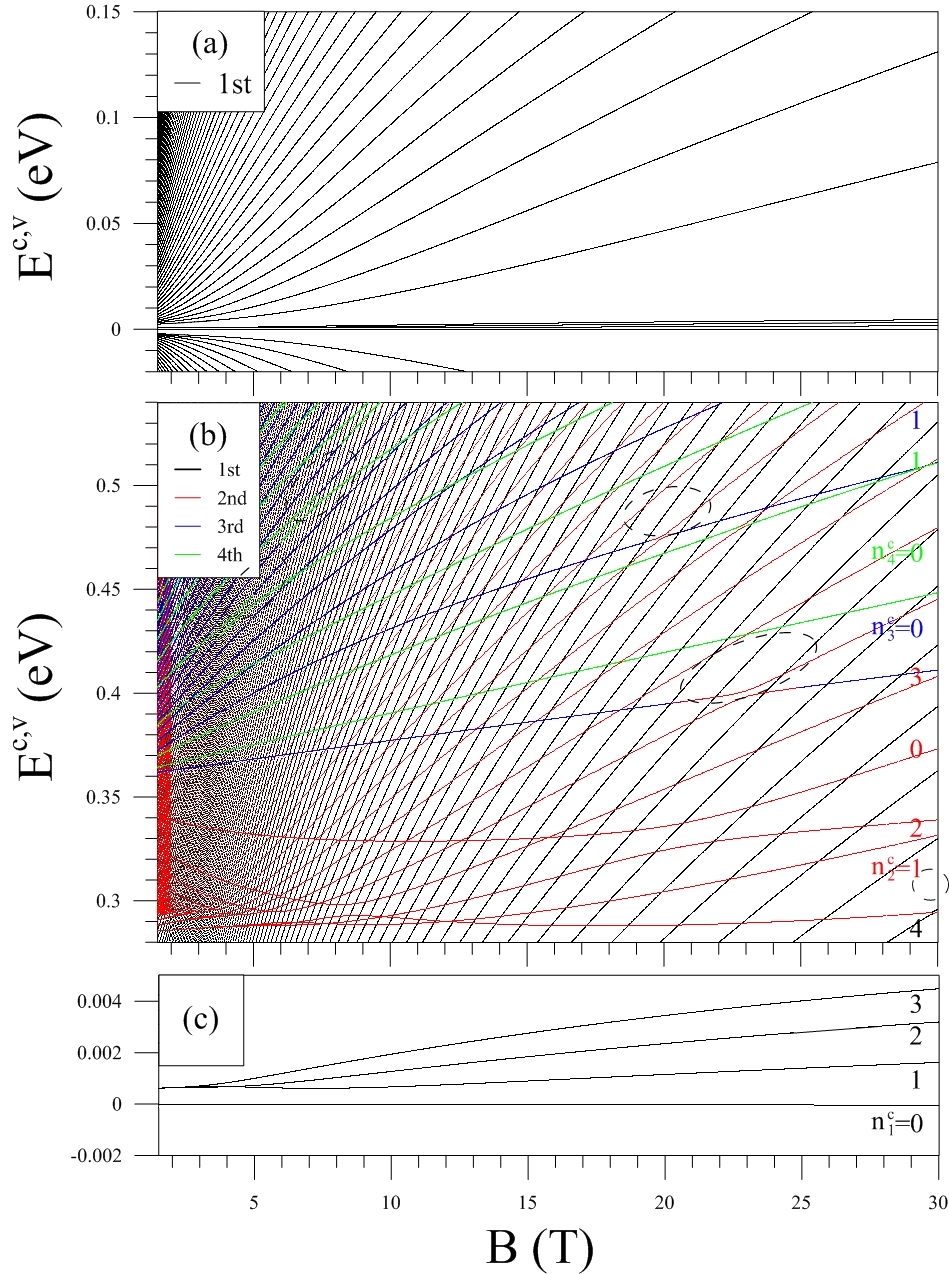}
    \caption{The B$_{z}$-dependent LL energy spectrum of the ABC-stacked
tetralayer graphene for (a) the first group, (b) the other three groups; (c)
the four LLs nearest to E$_{F}$.}
    \label{figure:4}
\end{figure}

\begin{figure}[p]
    \centering
    \includegraphics[width=1.0\textwidth]{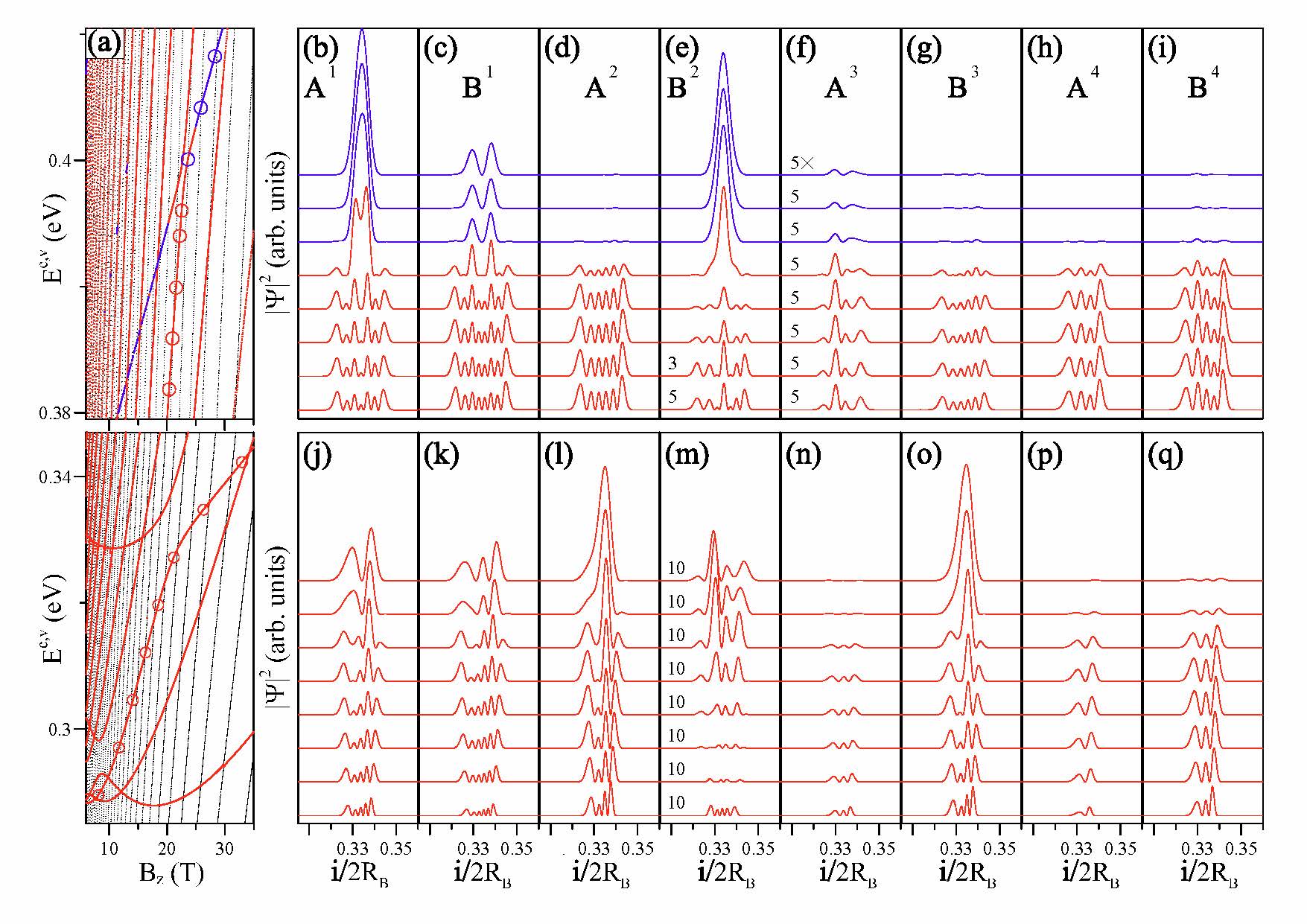}
    \caption{(a) The intragroup and intergroup LL anticrossing phenomena in the
ABC-stacked tetralayer graphene. The spatial evolutions of subenvelope
functions are shown in (b)-(i) for the second group of LLs, and (j)-(q)
for the second and the third groups of LLs.}
    \label{figure:5}
\end{figure}

\begin{figure}[p]
    \centering
    \includegraphics[width=0.9\textwidth]{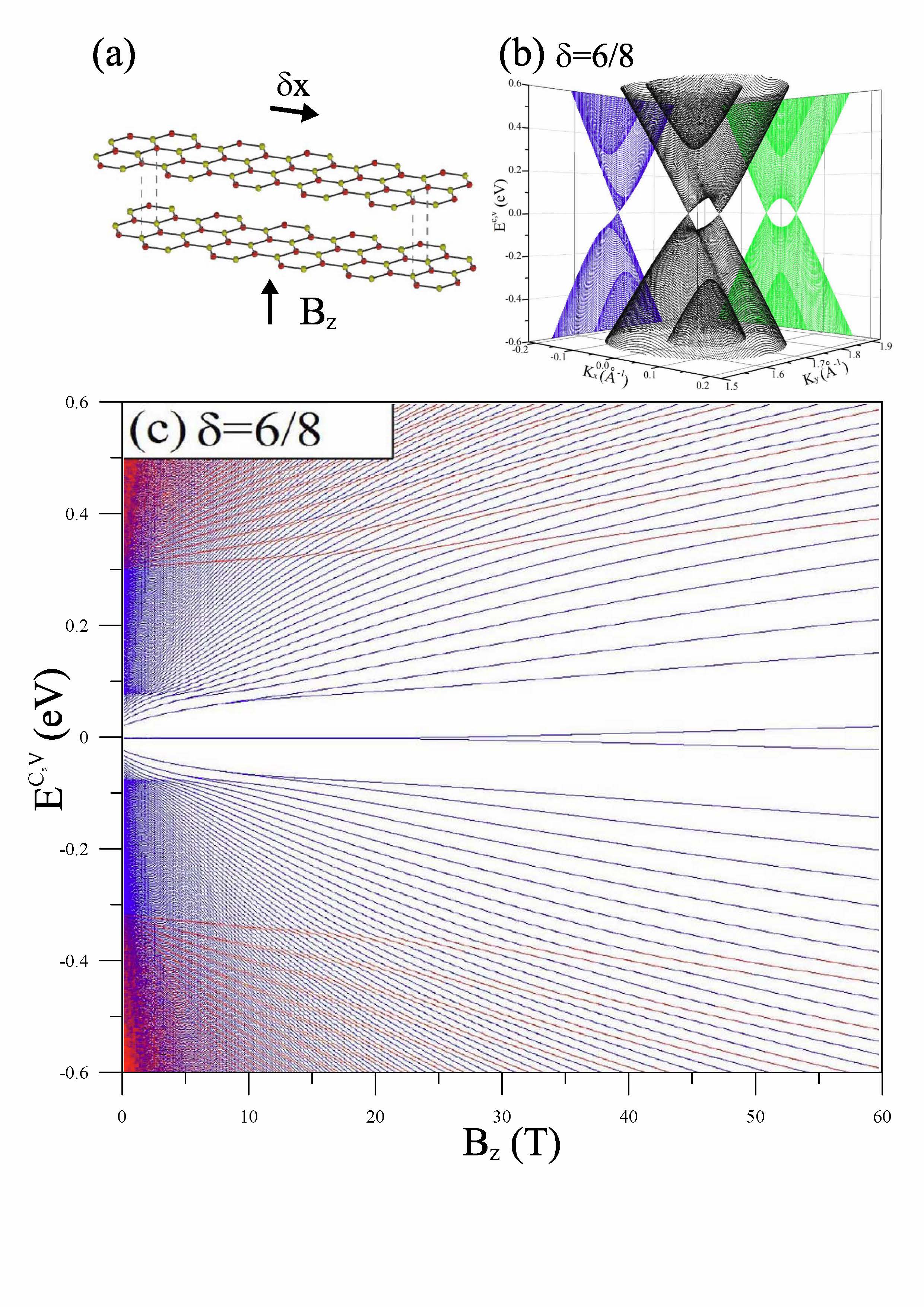}
    \caption{(a) Geometric structure of sliding bilayer graphene along the
armchair direction; (b) energy bands and (c) B$_{z}$-dependent LL spectrum
at $\delta $=6b/8.}
    \label{figure:6}
\end{figure}

\begin{figure}[p]
    \centering
    \includegraphics[width=0.8\textwidth]{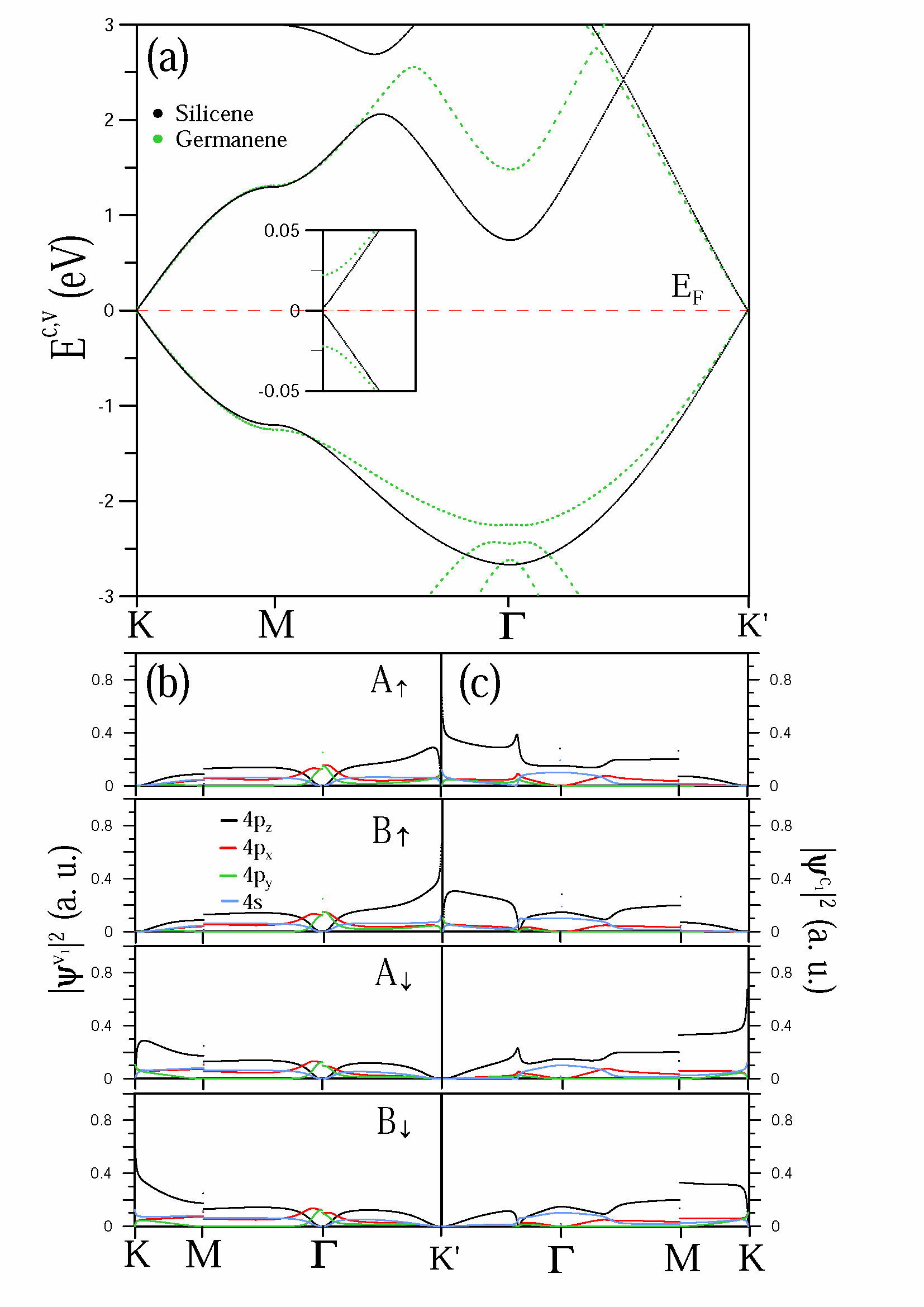}
    \caption{(a) Energy bands of monolayer germanene and silicene, and (b)\&(c)
orbital-decomposed state probabilities along the high-symmetry points. The
inset of (a) is the band structure near the Dirac point.}
    \label{figure:7}
\end{figure}

\begin{figure}[p]
    \centering
    \includegraphics[width=0.85\textwidth,trim=0 100 0 70, clip]{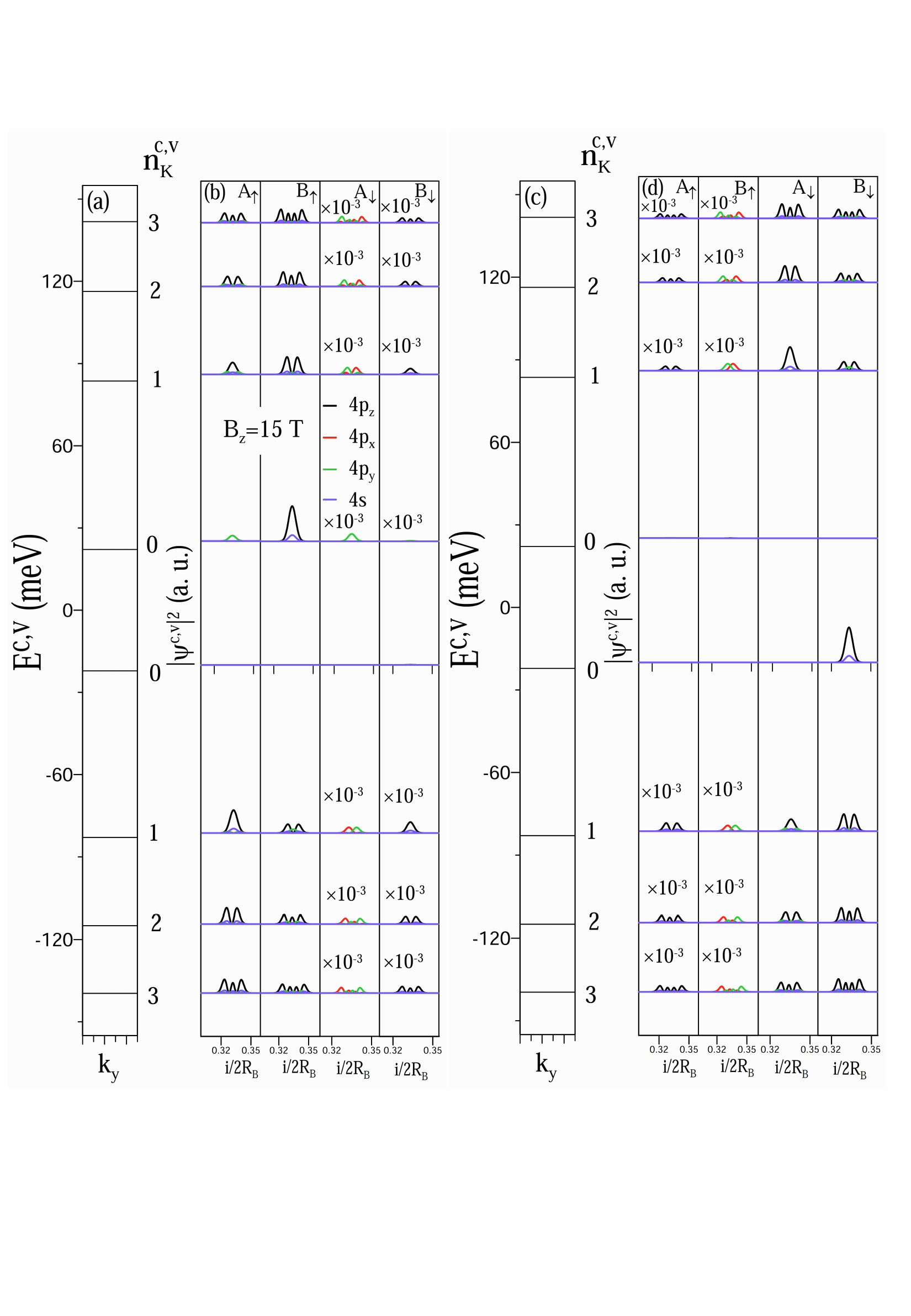}
    \caption{For germanene at B$_{z}$= 15 T, (a) \& (b) the up-dominated and
(c) \&(d) the down-dominated LL energies and spatial probability
distributions, corresponding to the quantized $K$-valley states.}
    \label{figure:8}
\end{figure}

\begin{figure}[p]
    \centering
    \includegraphics[width=0.8\textwidth]{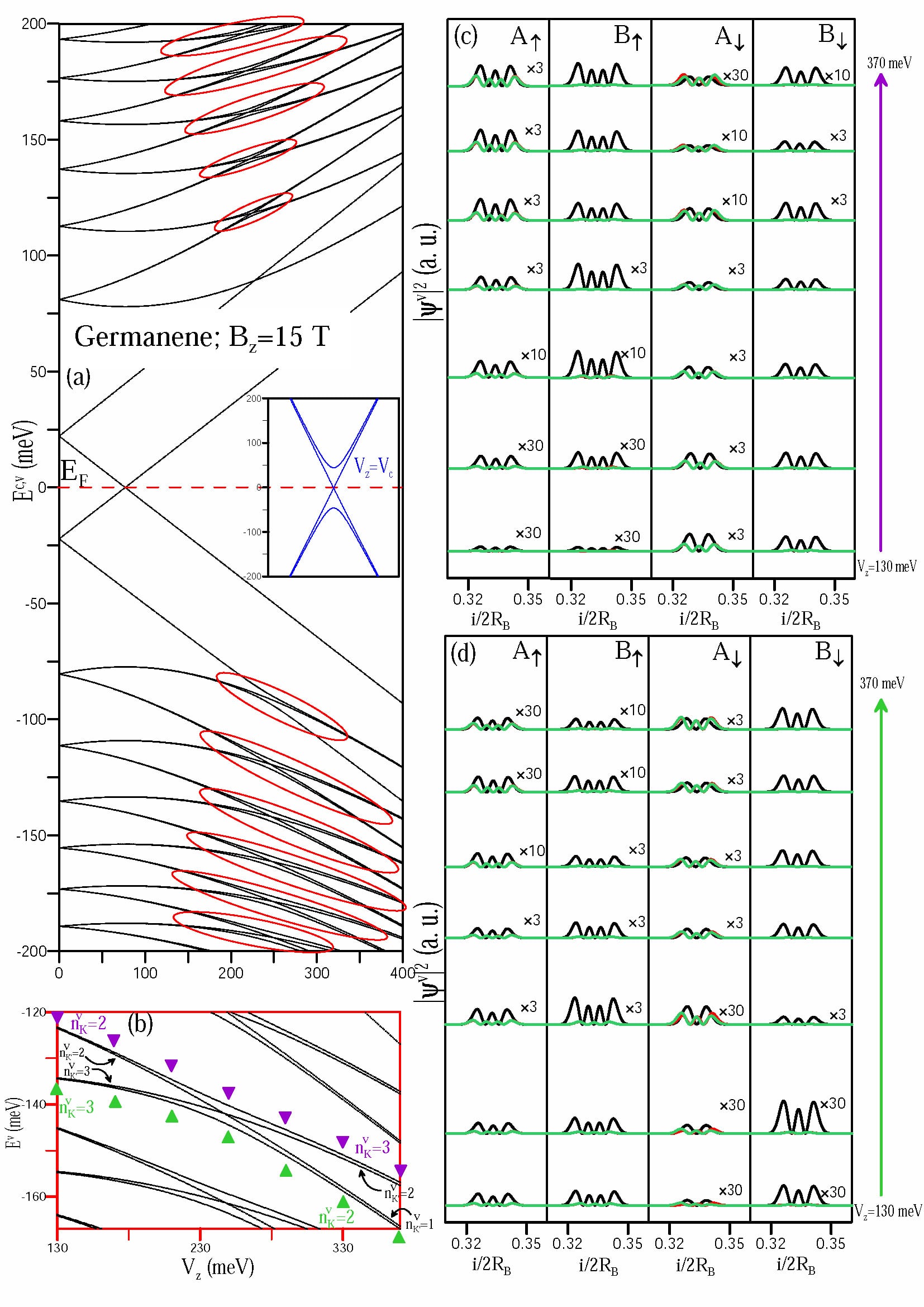}
    \caption{(a) The V$_{z}$-dependent LL energy spectra of germanene at B$_{z}$%
=15 T, and (b) the LL crossing and anticrossing within a certain range of E$%
^{v}$. (c) and (d) the drastic changes of probability distributions during
the LL anticrossings. The inset of (a) is the band structure at B$_{z}$=0
and a critical V$_{z}$. The LL anticrossings are indicated by the red
circles.}
    \label{figure:9}
\end{figure}

\begin{figure}[p]
    \centering
    \includegraphics[width=1.0\textwidth,trim=0 200 0 100, clip]{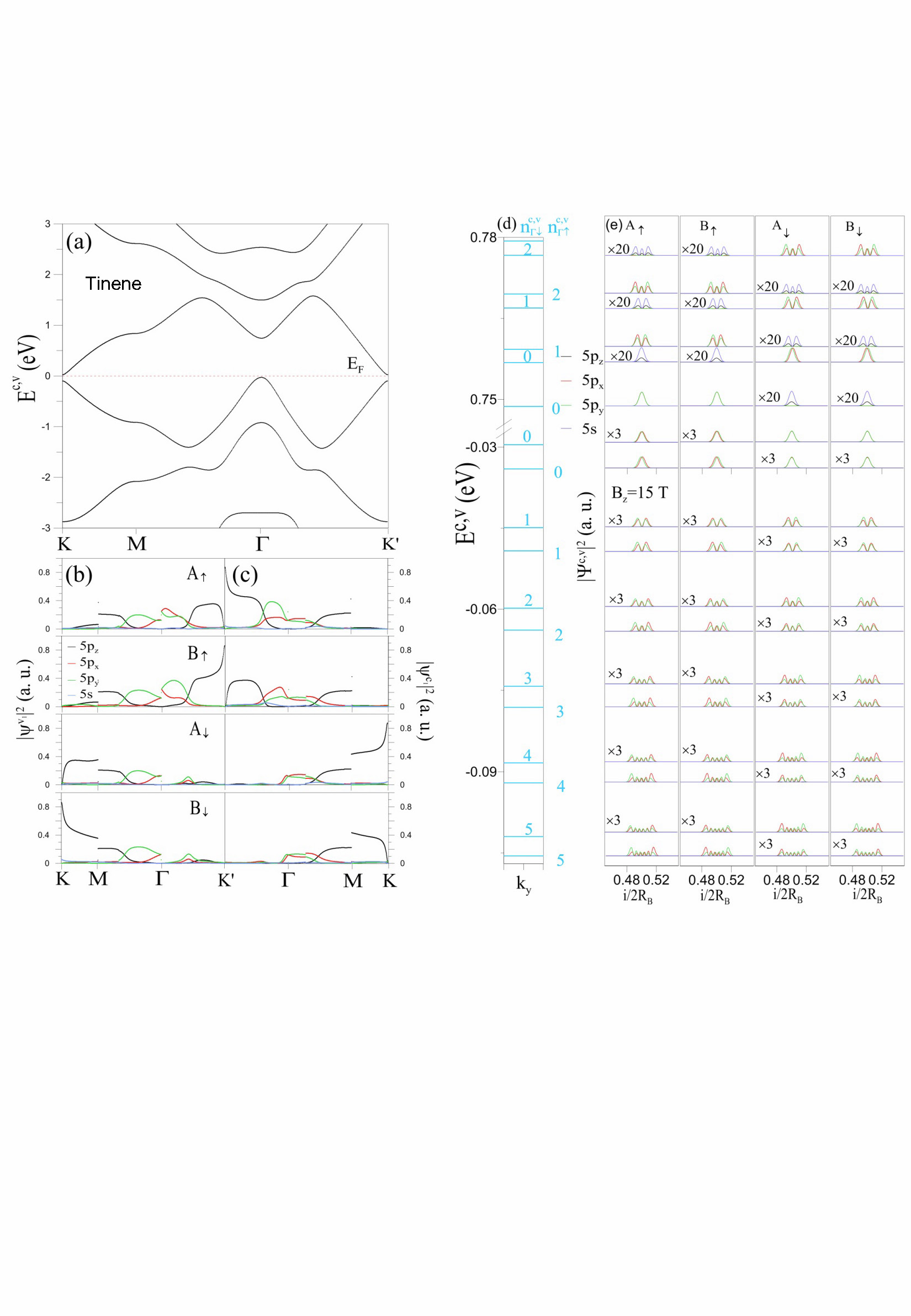}
    \caption{(a)-(c) Similar plot as Fig. 7, but shown for monolayer tinene.
Also plotted in (b) are those from the quantized $\Gamma $-valley states.}
    \label{figure:10}
\end{figure}

\begin{figure}[p]
    \centering
    \includegraphics[width=0.8\textwidth]{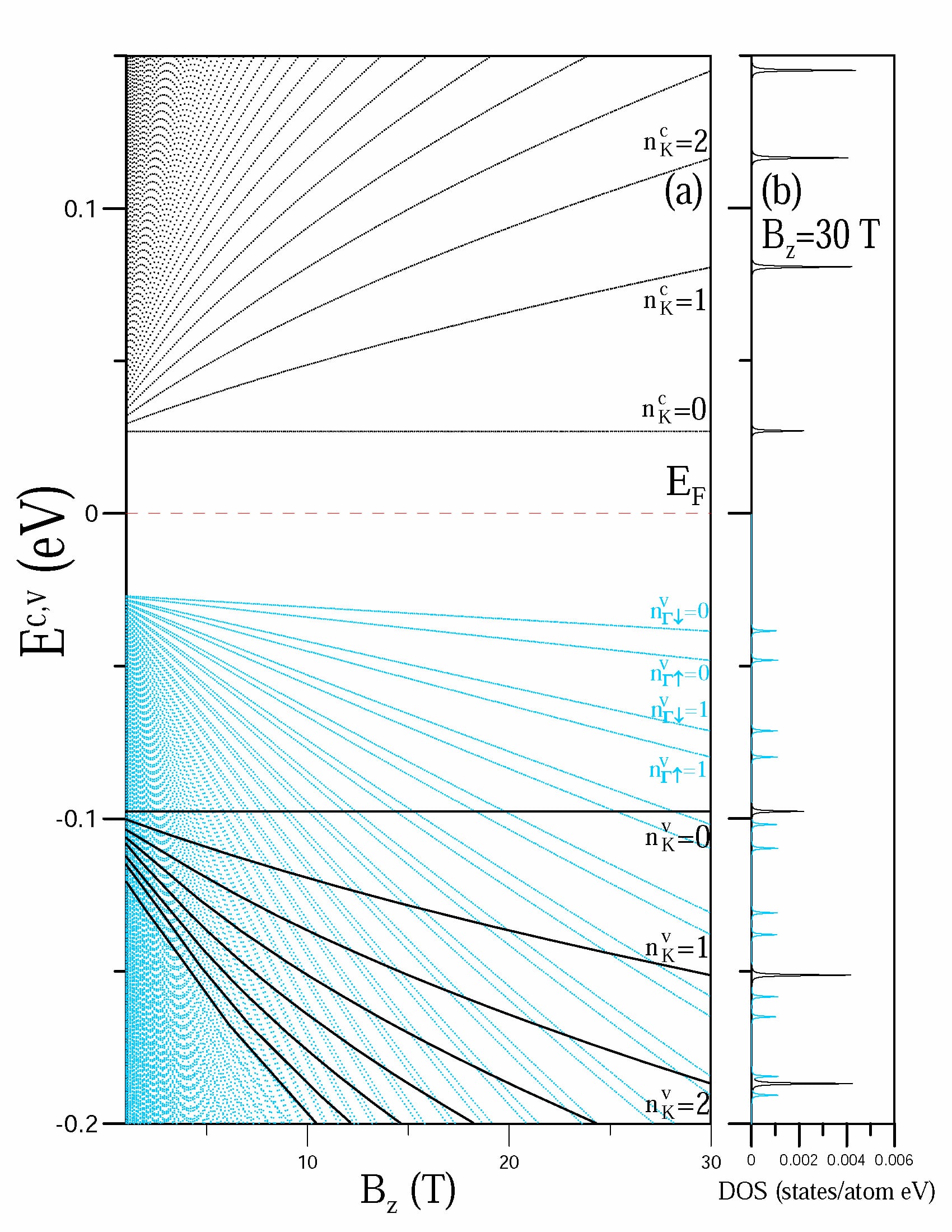}
    \caption{(a) The B$_{z}$-dependent energy spectra of the first and second
groups (black and blue curves) in tinene, and (b) density of states at B$%
_{z} $ = 30 T.}
    \label{figure:11}
\end{figure}

\begin{figure}[p]
    \centering
    \includegraphics[width=0.9\textwidth]{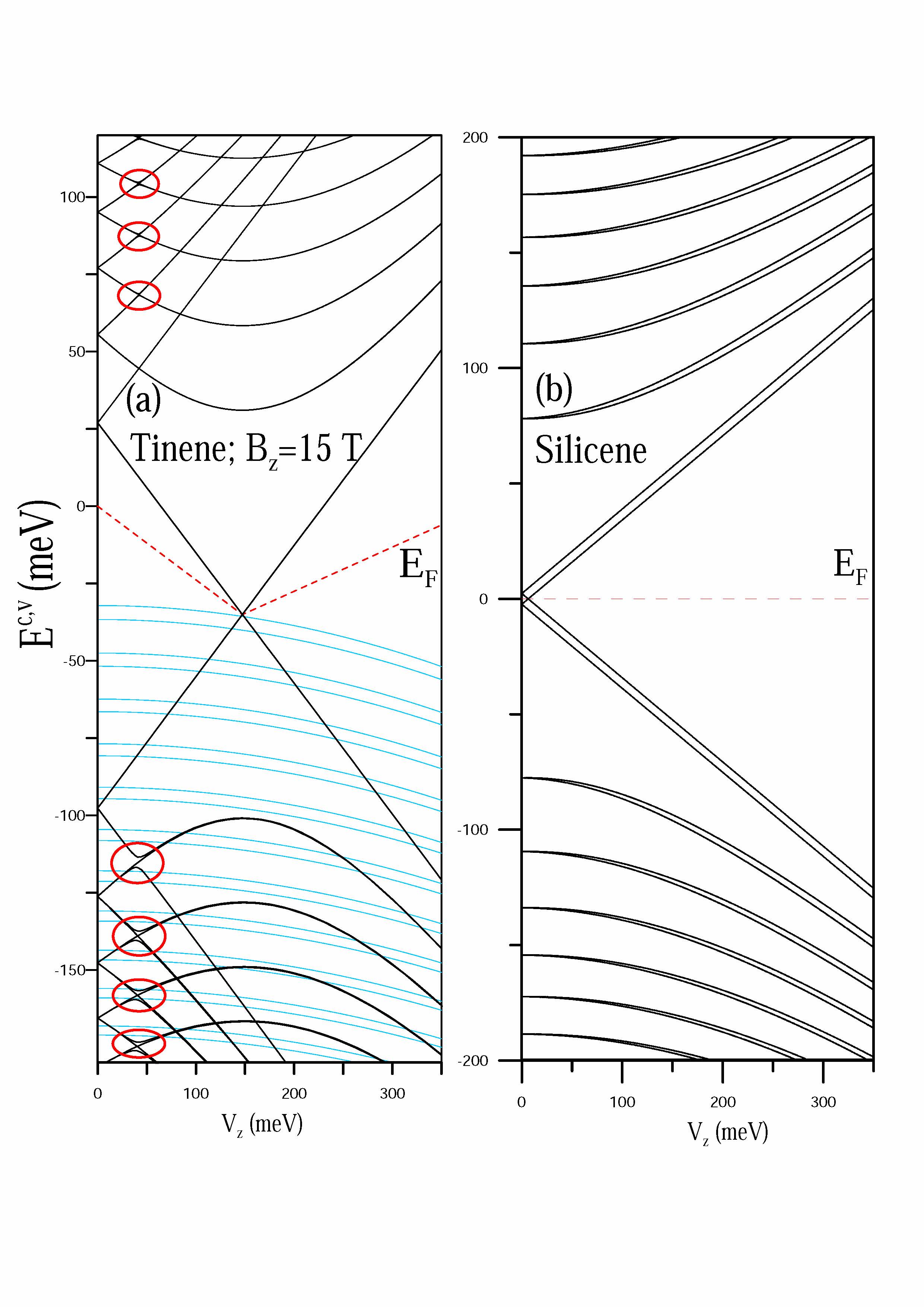}
    \caption{The V$_{z}$-dependent LL energy spectra at B$_{z}$=15 T for (a)
tinene and (b) silicene.}
    \label{figure:12}
\end{figure}

\begin{figure}[p]
    \centering
    \includegraphics[width=0.8\textwidth]{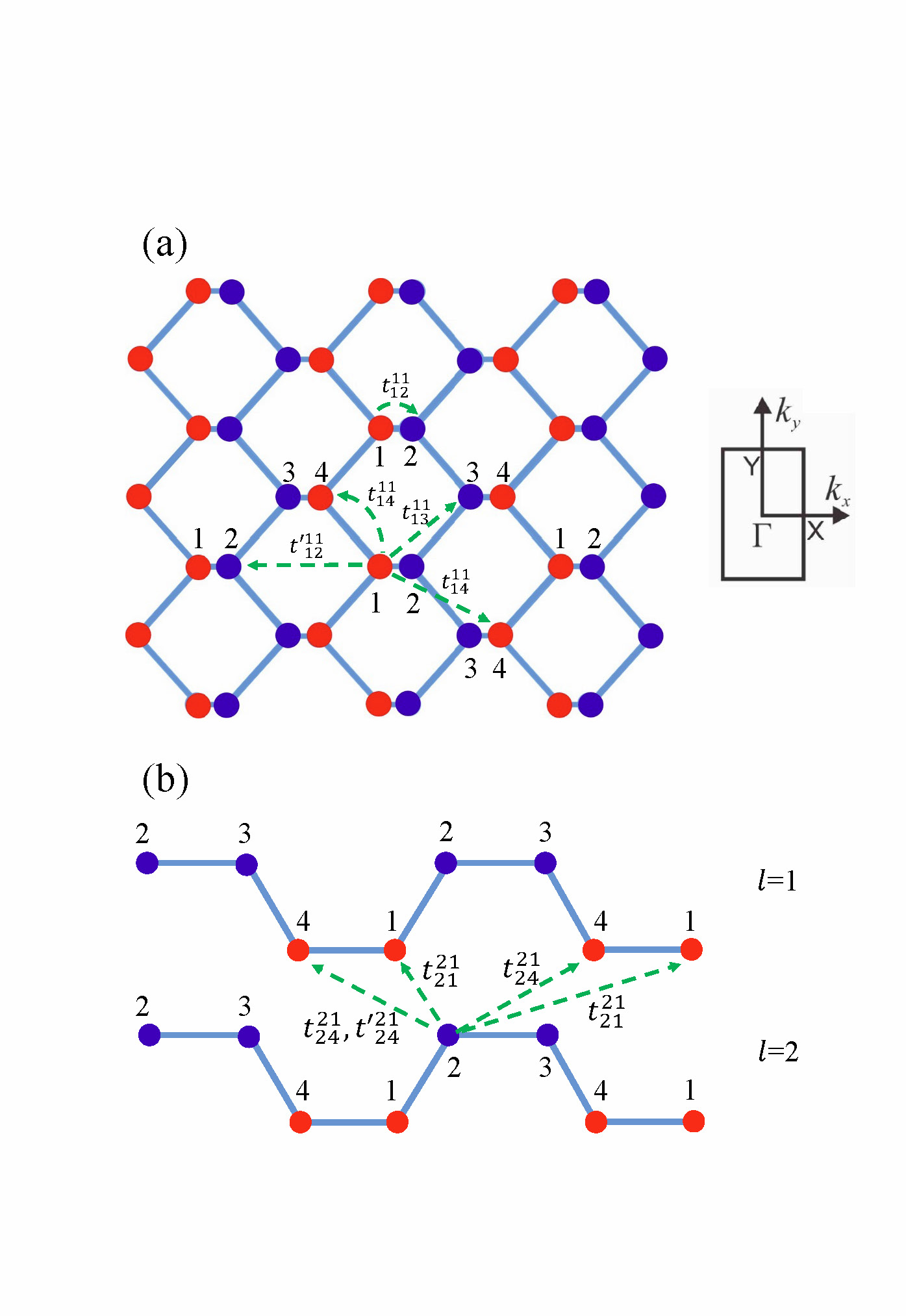}
    \caption{(a) The geometric structure and the first Brillouin zone of
monolayer phosphorene, and (b) the AB stacking structure of bilayer
phosphorene. Also, the intralayer and interlayer hopping integrals are
marked in (a) and (b), respectively.}
    \label{figure:13}
\end{figure}

\begin{figure}[p]
    \centering
    \includegraphics[width=0.8\textwidth]{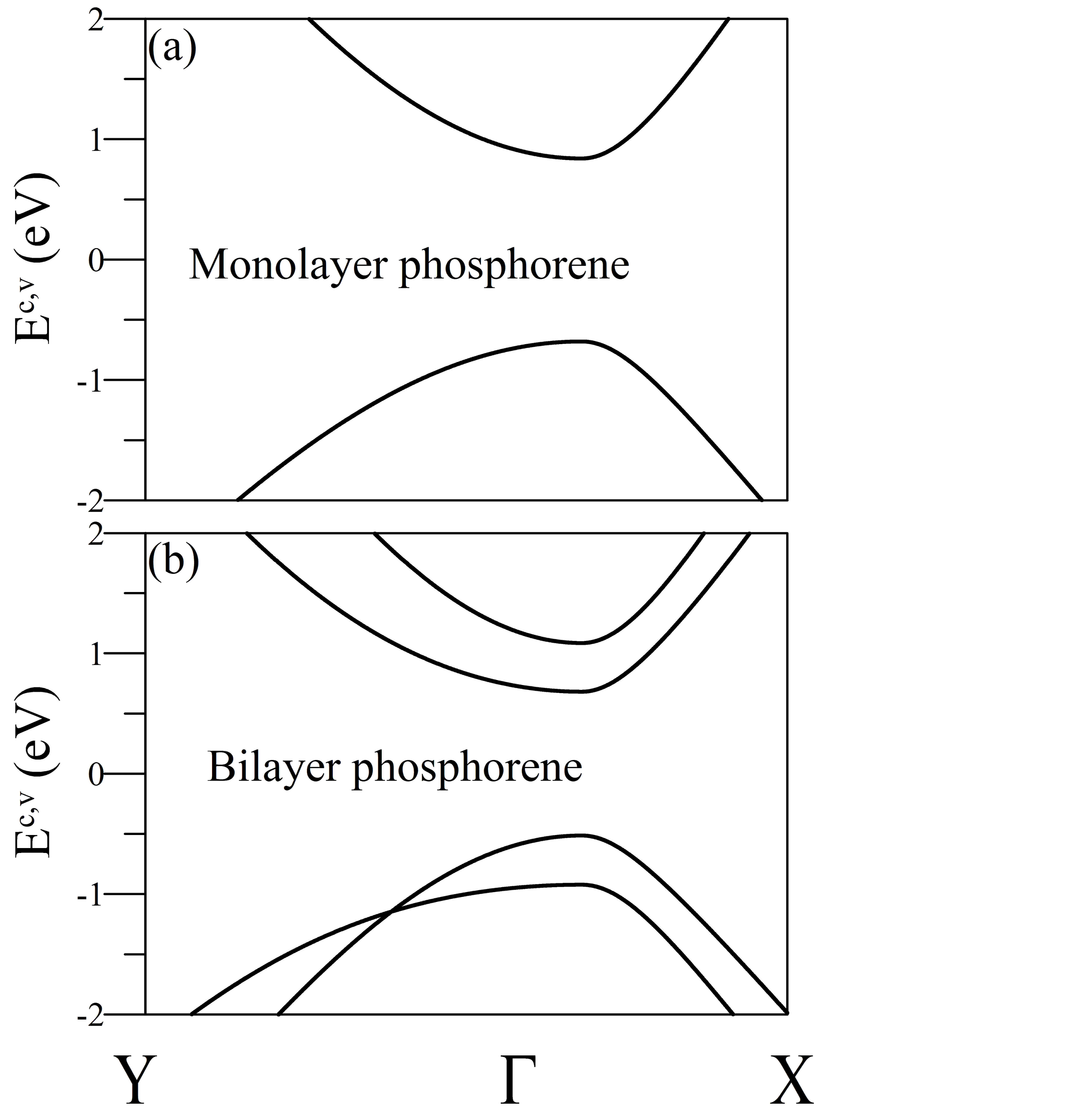}
    \caption{Energy bands of (a) monolayer and (b) bilayer phosphorene.}
    \label{figure:14}
\end{figure}

\begin{figure}[p]
    \centering
    \includegraphics[width=0.8\textwidth]{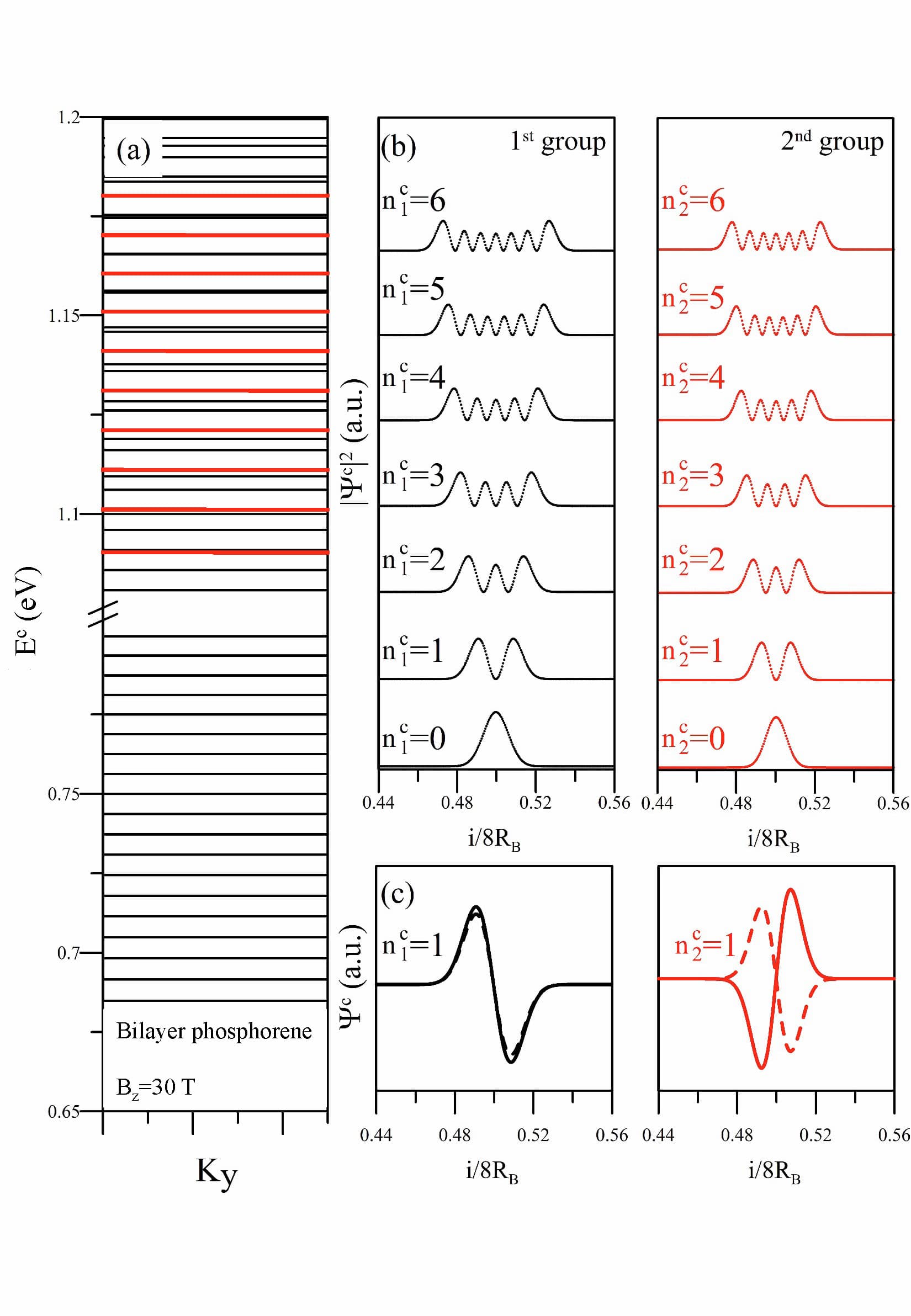}
    \caption{(a) The LL energies of bilayer phosphorene at B$_{z}$=30 T and
(b) the probability distributions. Also shown in (c) are the amplitudes of $%
n^{c}=1$ of the upper (solid curve) and lower (dashed curve) layers.}
    \label{figure:15}
\end{figure}

\begin{figure}[p]
    \centering
    \includegraphics[width=0.8\textwidth]{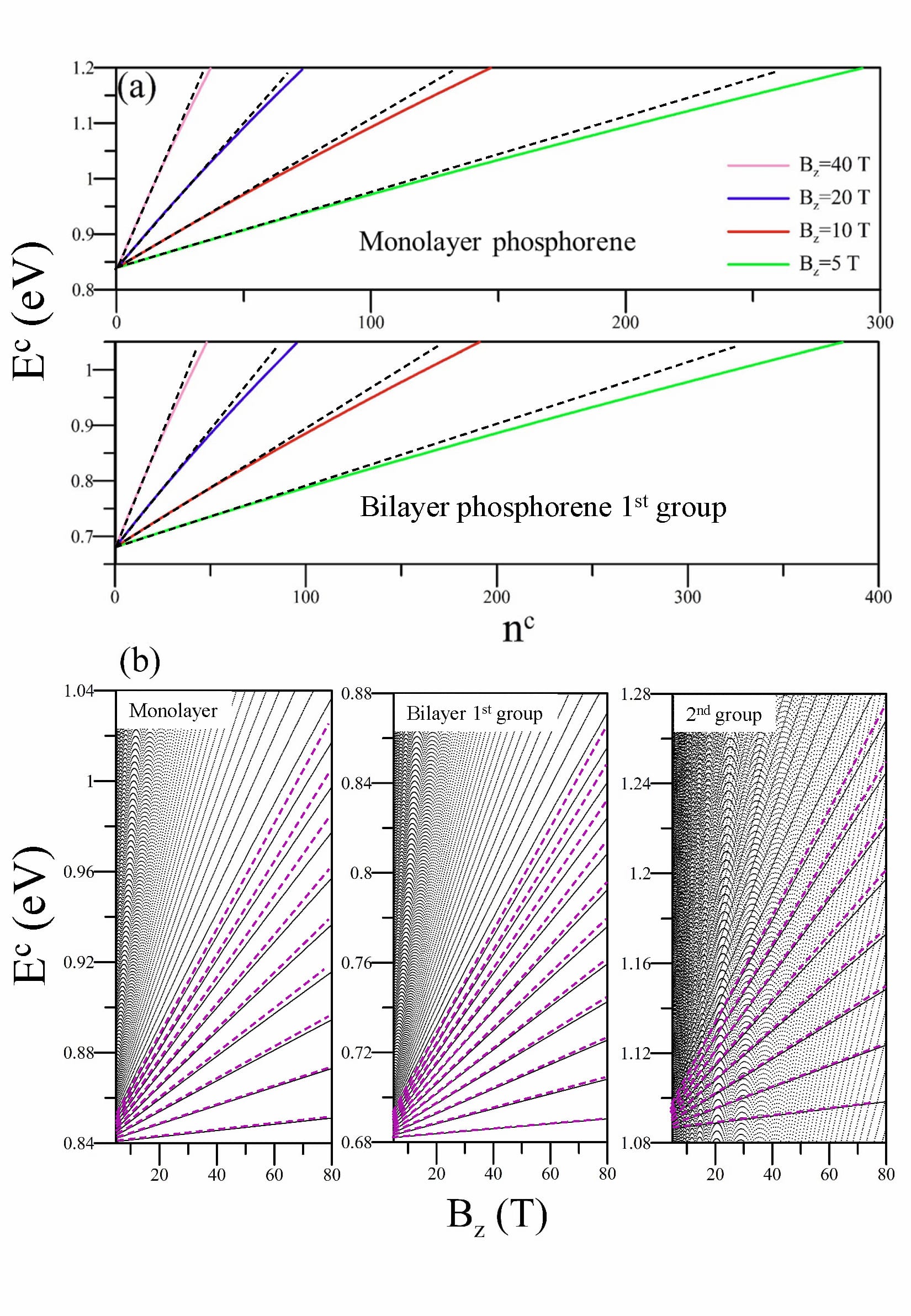}
    \caption{For monolayer and bilayer phosphorene, the $n^{c}$- and $B_{z}$%
-dependent LL energies are shown in (a) and (b) respectively. The black and
purple dashed lines in (a) and (b) represent the linear dependence.}
    \label{figure:16}
\end{figure}

\begin{figure}[p]
    \centering
    \includegraphics[width=1.0\textwidth,trim=0 200 0 0, clip]{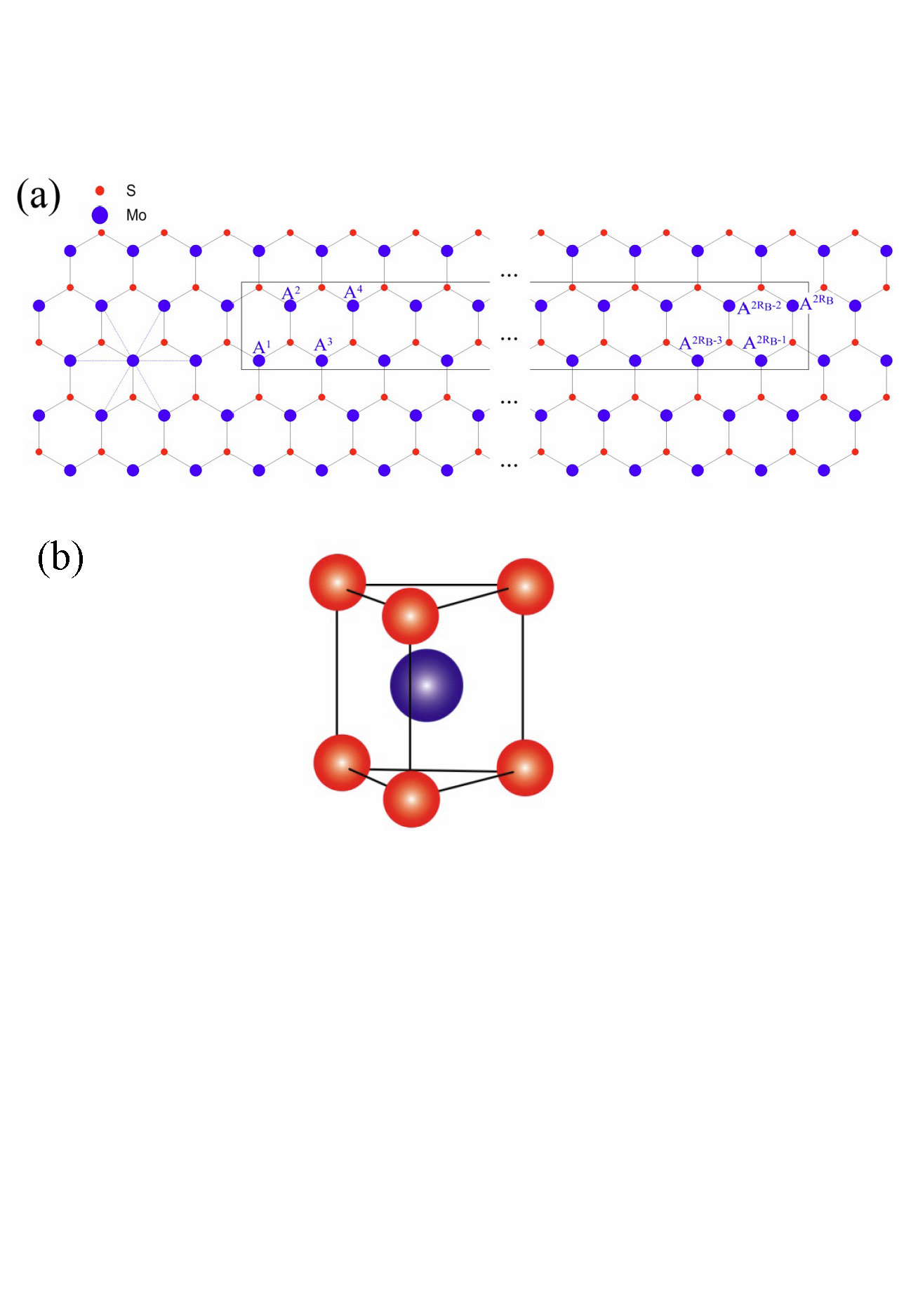}
    \caption{(a) Geometric structures for MoS$_{2}$ monolayer with an enlarged
rectangular unit cell in B$_{z}\widehat{z}$ and (b) the structure of
trigonal prismatic coordination.}
    \label{figure:17}
\end{figure}

\begin{figure}[p]
    \centering
    \includegraphics[width=0.7\textwidth]{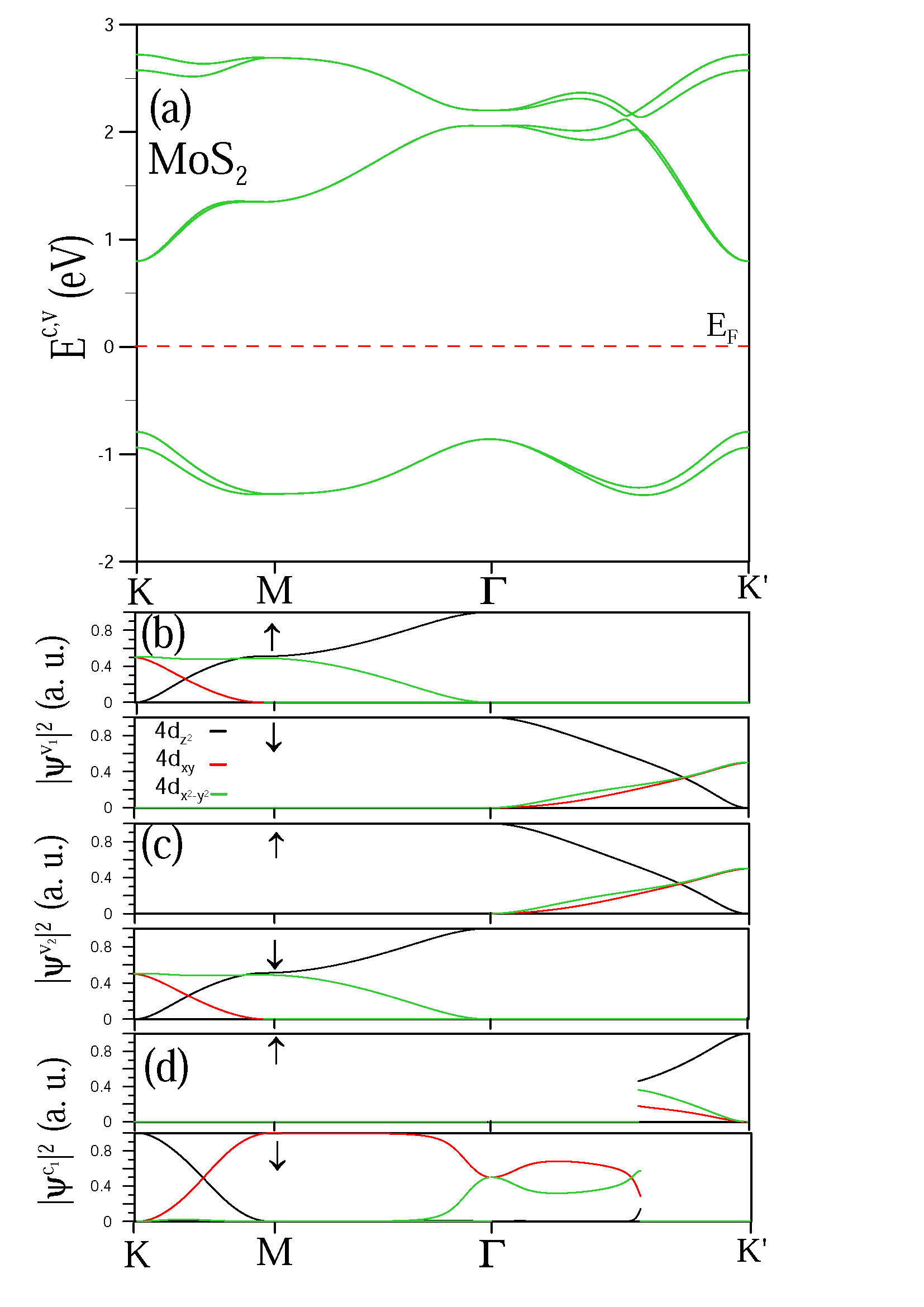}
    \caption{(a) Energy bands of monolayer MoS$_{2}$, and (b)-(d) the
orbital-decomposed state probabilities along the high-symmetry points.}
    \label{figure:18}
\end{figure}

\FloatBarrier

\begin{figure}[p]
    \centering
    \includegraphics[width=0.7\textwidth]{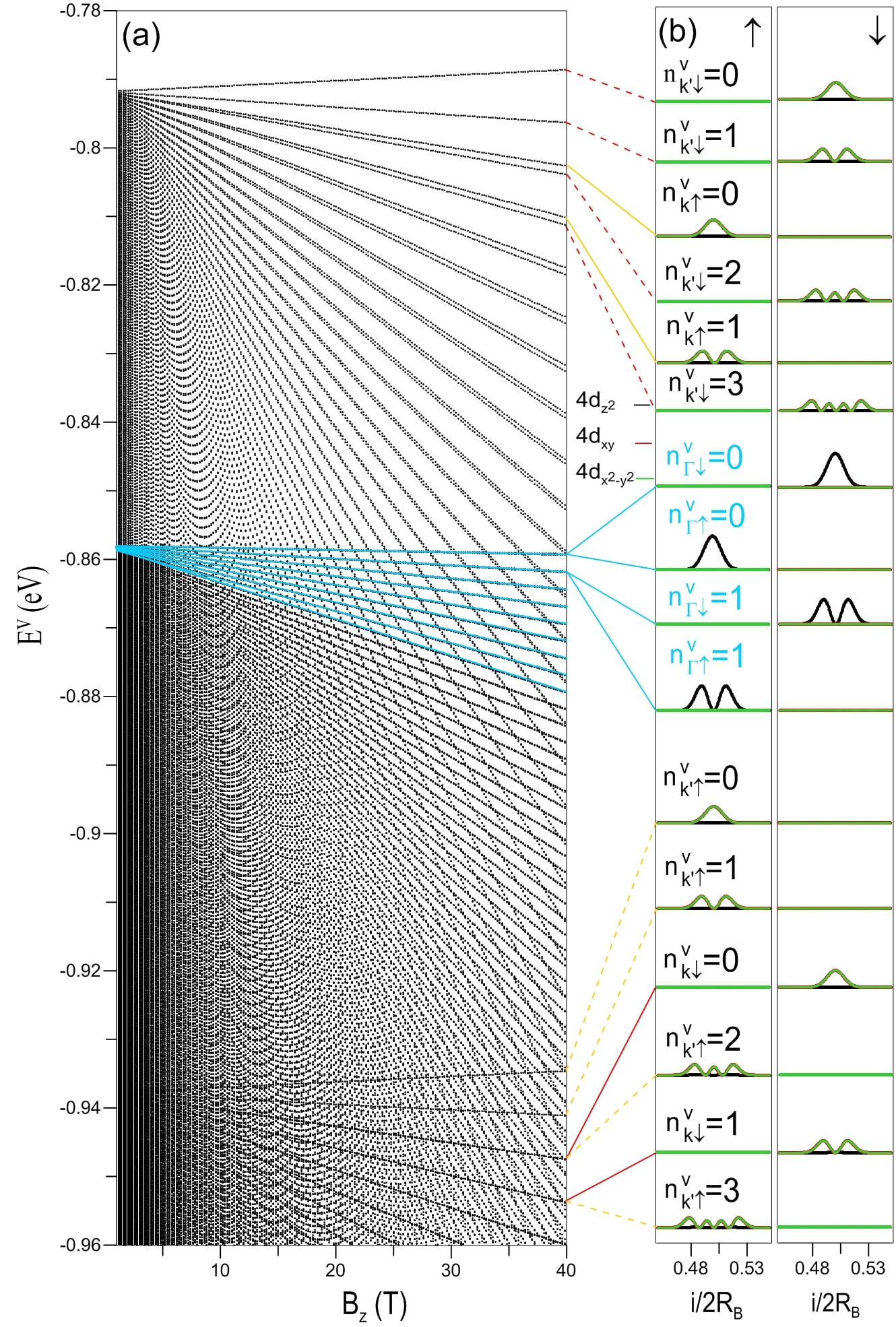}
    \caption{(a) The B$_{z}$-dependent energy spectra of the valence LLs are,
respectively, related to the quantized states near the ($K$,$K^{^{\prime }}$%
) and $\Gamma $ points (black and blue curves), in which the
valley-dependent (spin-dependent) subgroups are represented by the solid and
dashed curves (the red and yellow colors). The spatial probability
distributions are shown in (b) at B$_{z}$=40 T.}
    \label{figure:19}
\end{figure}

\begin{figure}[p]
    \centering
    \includegraphics[width=0.8\textwidth]{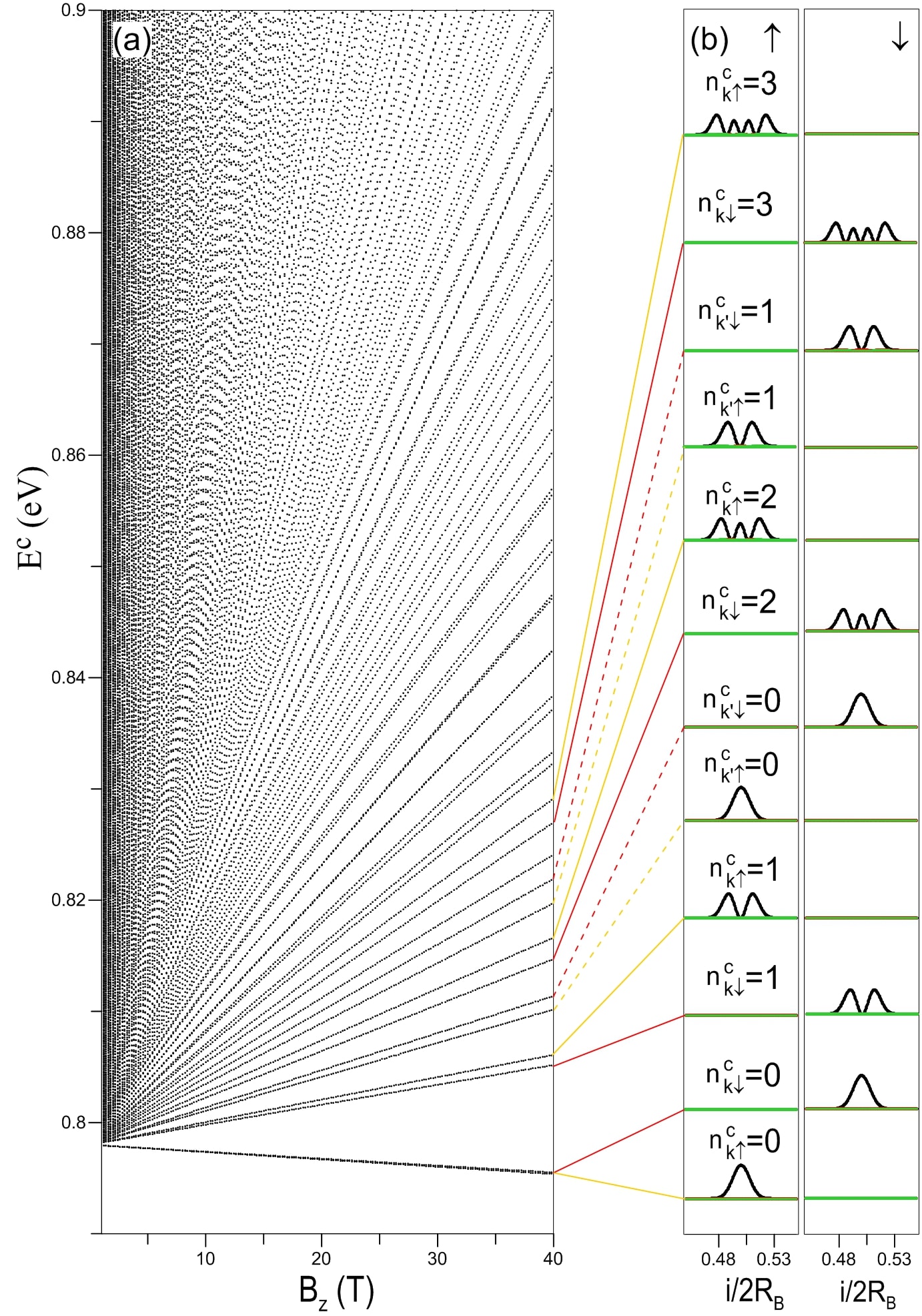}
    \caption{Same plot as Fig. 19, but shown for the low-lying conduction LLs.}
    \label{figure:20}
\end{figure}

\end{document}